\documentclass[%
reprint
,twocolumn%
,nobibnotes,pre,floatfix,superscriptaddress]{revtex4-2}
\usepackage{amsthm}
\usepackage{ulem}
\usepackage{graphicx}
\usepackage{amsfonts}
\usepackage{amsmath}
\usepackage{bbold}
\usepackage{bm}
\usepackage{enumerate}
\usepackage{color}
\newcommand{\uvec}[1]{{\boldsymbol{\hat{#1}}}}
\newcommand{\bvec}[1]{\boldsymbol{#1}}

\begin{document}
\title{Transition to Synchrony in a Three-Dimensional Swarming Model with Helical Trajectories}
\author{Chunming Zheng}
\author{Ralf Toenjes}
\affiliation{Institute for Physics and Astronomy, University of Potsdam, Karl-Liebknecht-Strasse 24/25, 14476 Potsdam-Golm, Germany}
\author{Arkady Pikovsky}
\affiliation{Institute for Physics and Astronomy, University of Potsdam, Karl-Liebknecht-Strasse 24/25, 14476 Potsdam-Golm, Germany}
\affiliation{Department of Control Theory, Nizhny Novgorod State University, Gagarin Avenue 23, 606950 Nizhny Novgorod, Russia}
\date{\today}
\begin{abstract}
We investigate the transition from incoherence to global collective motion in a three dimensional swarming model of agents with helical trajectories, subject to noise and global coupling. Without noise this model was recently proposed as a generalization of the Kuramoto model and it was found, that alignment of the velocities occurs discontinuously for arbitrary small attractive coupling. Adding noise to the system resolves this singular limit and leads to a continuous transition, either to a directed collective motion, or to center of mass rotations.
\end{abstract}
\maketitle
\section{Introduction}
\label{sec:Introduction}

Helical motion is a common form of movement in active particles, e.g, micro-swimmers using flagella for propulsion \cite{lauga2009hydrodynamics,bechinger2016active}. It facilitates chemotaxis even for small particles. Oscillating in circles much larger than the body size, biological swarmers can detect chemical gradients and adapt their translational motion accordingly. Moreover, artificial swarmers, such as  magnetic micromachines with helical motion \cite{tottori2012magnetic} or 
microrobot swarms \cite{xie2019reconfigurable} are being designed and controlled in the lab  with potential biomedical applications, e.g., for drug delivery. When such self-propelled particles interact, their velocities can align resulting in a directed collective motion \cite{vicsek1995novel,attanasi2014finite,chen2017weak}. In addition to a directional alignment, phase  synchronization of oscillatory movements may also be possible, resulting in collective oscillations. 

 The seminal Vicsek model~\cite{vicsek1995novel} of swarming particles, despite its simple formulation, displays a variety of dynamical regimes \cite{gregoire2004onset,chate2008collective}. Its basic approximation is that active particles in a viscous medium move at velocities $\uvec{v}$ with constant (unit) amplitude and only adjust their directions through interactions with neighboring particles. The Vicsek model can  easily be extended to include helical trajectories by defining individual rotation axes $\uvec{\omega}$ 
 and frequencies (angular velocities)  $\omega$ for the particle velocity vectors. In general, the velocities and the rotation directions evolve in time, are coupled, and are subject to noise. As an ubiquitous influence in nature, noise plays an important, often antagonizing, role in the dynamics of the collective motion, in particular at microscopic scales.
 
Without noise, and with a fixed distribution of frequencies and static rotation axes, 
this setup has recently been proposed and analyzed as a high-dimensional 
generalization of the Kuramoto model \cite{chandra2019continuous}. It was found, that 
for odd-dimensional vectors $\uvec{v}$, the synchronization transition occurs 
discontinuously and without hysteresis for arbitrarily small attractive coupling. 
This means that 
in three dimensions frequency heterogeneity cannot prevent synchronization at 
small coupling strengths. We report in this paper 
that this is the singular limit of a transition at 
finite coupling strength in the presence of noise.  The Watanabe-Strogatz 
theory \cite{watanabe1994constants} and the Ott-Antonsen ansatz \cite{ott2008low},
first developed for 
ensembles of two-dimensional noise-free oscillators, have been shown to generalize to higher dimensions \cite{tanaka2014solvable,chandra2019complexity} as well. 
With noise, identical frequencies and certain fixed distributions of rotation axes, the stability of the incoherent (uniform) velocity distribution has been obtained for an equivalent system of random tops \cite{ritort1998solvable}, a mechanical model for a disordered spin system. The magnetization transition in this model corresponds to a directed collective motion in the swarming model. In another context, a spatio-temporal alignment of vectors rotating on a unit sphere may also be considered a very simplified model for beating cilia, which in general rotate under a variable angle around a fixed axis \cite{nieder2008}. 

In this paper we present a general condition for the transition to collective motion (alignment) for arbitrary but fixed distributions of rotation axes and heterogeneous frequencies, based on a linear stability analysis. This condition can still be used in an adiabatic approximation if the rotation axes $\uvec{\omega}$ are not fixed but evolve on a longer timescale than the particle velocities $\uvec{v}$. In this case the stability of the incoherent state depends adiabatically on the degree of the rotation axes alignment.

\section{Model formulation}

\subsection{Langevin equation}
Independent of their interpretation as velocities, we are considering 
a set of $N$ unit vectors $\uvec{\sigma}_i$ with $i=1\dots N$, 
subject to torques $\bvec{\mu}_i$
\begin{equation}\label{Eq:SphereForce}
	\dot{\uvec{\sigma}}_i = \bvec{\mu}_i\times\uvec{\sigma}_i\;.
\end{equation}
The forces act perpendicular to the vectors $\uvec{\sigma}_i$, ensuring that the amplitudes remain constant. Throughout the text we denote vectors by bold symbols and mark unit vectors, such as  $\uvec{\sigma}$, with hats. Symbols subscripted with $x$, $y$ and $z$ denote vector components in cartesian coordinates. The torque $\bvec{\mu}_i$ can be any time-dependent global or individual forcing. We assume it to be the sum of three components: (i)  a constant rotation bias of amplitude $\omega_i$ around a rotation axis in the direction $\uvec{\omega}_i$; (ii) an alignment force which rotates $\uvec{\sigma}_i$ towards a vector $\bvec{\rho}_i$ (this component is responsible for interaction of the units);  and (iii) a noise component $\bvec{\xi}_i$:
\begin{equation}\label{Eq:Torque}
	\bvec{\mu}_i = \omega_i\uvec{\omega}_i + K \left(\uvec{\sigma}_i\times\bvec{\rho}_i\right)+\bvec{\xi}_i(t).
\end{equation}
Here $K$ is a coupling strength which, when it is positive, promotes alignment of $\uvec{\sigma}_i$ with $\bvec{\rho}_i$.  The term $\bvec{\xi}_i(t)$ is a vector of independent Gaussian white noises $\left\langle (\bvec{\xi}_i)_n(t)(\bvec{\xi}_j)_m(t')\right\rangle = 2D\delta_{ij}\delta_{mn}\delta(t-t')$. The Langevin equation \eqref{Eq:SphereForce} with stochastic force \eqref{Eq:Torque} is to be interpreted in the sense of Stratonovich to preserve the unit amplitude of the vectors $\uvec{\sigma}_i$. By direct simulation of the model we observe that a positive global coupling above some critical value leads to an alignment (synchronization) of the units, as shown in Figs.~\ref{Fig:Rho_vs_DK},\ref{Fig:Examples}. The goal of the analysis below is to understand this transition.

\subsection{Fokker-Planck equation}

In the standard Vicsek model~\cite{vicsek1995novel} with local interactions, 
the variables $\uvec{\sigma}_i$ are particle velocities $\uvec{v}_i$, the constant rotation bias is zero ($\omega_i\uvec{\omega}_i=0$) and the vector $\bvec{\rho}_i$ is the average velocity of all particles within a distance $R$ from the $i$-th particle. As a result of the competition between the aligning coupling and noise, there exists a critical coupling strength $K_{cr}$, at which the incoherent state loses stability. When the radius of interaction is taken to be larger than the spatial size of the population, the coupling becomes global, i.e. $\bvec{\rho}_i =\bvec{\rho} = \frac{1}{N}\sum_i \uvec{\sigma}_i = \langle\uvec{\sigma}\rangle$. Below we consider globally coupled populations only. The amplitude $\rho = |\bvec{\rho}|$ serves as the order 
parameter for the synchronization/alignment transition. It takes values between zero for a uniform distribution of $\uvec{\sigma}_i$ and one when the vectors are identical.

In the thermodynamic limit $N\to\infty$ the system can be described by a family of smooth densities $f(\uvec{\sigma},t;\uvec{\omega},\omega)$ 
for vectors $\uvec{\sigma}$ with a given fixed rotation bias $\omega\uvec{\omega}$. These densities obey the Fokker-Planck 
equation
\begin{equation}
	\partial_t f + \bvec{\nabla}_s\cdot(f\bvec{a})=D\bvec{\nabla}^2_sf,
\label{Eq:FP}
\end{equation}
where $\bvec{\nabla}_s$ is the vector differential operator along the surface of a unit sphere acting on the argument $\uvec{\sigma}$ and
$\bvec{a} = \omega\uvec{\omega}\times\uvec{\sigma} + K\left(\bvec{\rho}-(\bvec{\rho}\cdot\uvec{\sigma})\uvec{\sigma}\right) $ is the deterministic part of the force acting on a vector $\uvec{\sigma}$ with rotation bias $\omega\uvec{\omega}$. 

In this paper we assume that the frequencies $\omega$ and the rotation axes $\uvec{\omega}$ are random and independent.
They are distributed according to the probability densities $g(\omega)$ and $G(\uvec{\omega})$, respectively. The order parameter $\bvec{\rho}$ is the expectation value
\begin{equation}\label{Eq:OrderParmDef}
	\bvec{\rho}=\int_{-\infty}^\infty d\omega~ g(\omega) \int_{S^2} \bvec{r}(\uvec{\omega},\omega) G(\uvec{\omega}) ~dA(\uvec{\omega})
\end{equation}
of the frequency dependent mean fields
\begin{equation}\label{Eq:FamilyOrder}
	\bvec{r}(\uvec{\omega},\omega) = \int_{S^2} \uvec{\sigma} f(\uvec{\sigma},t;\uvec{\omega},\omega)~  dA(\uvec{\sigma}).
\end{equation}
The terms $dA$ denote the $S^2$ surface volume elements.
\section{Diffusion on a sphere with global coupling}
\label{sec:GlobalVicsek}
\begin{figure}[t!]
\includegraphics[height=6cm]{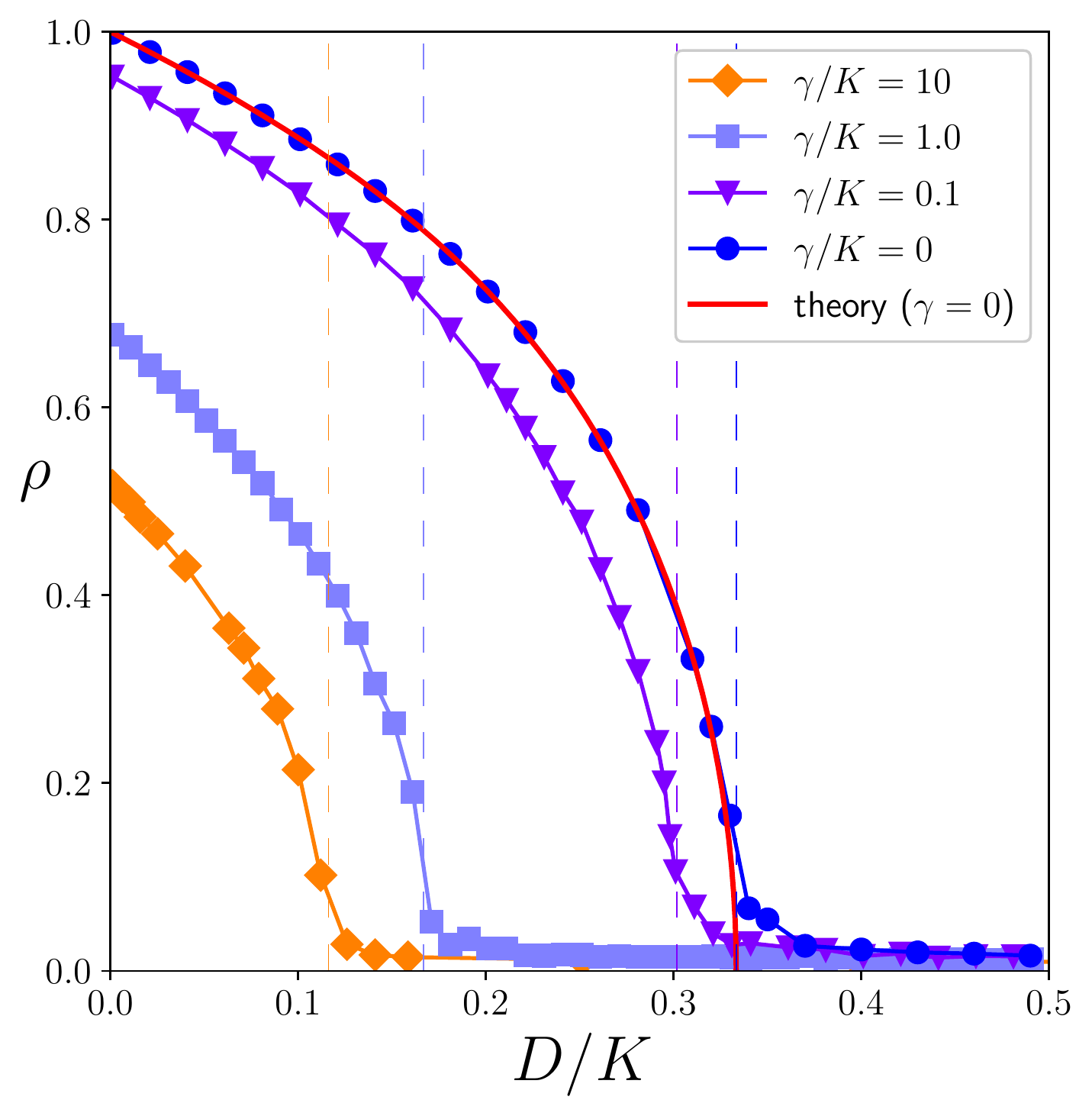}
\caption{Amplitudes of the stationary mean velocity for particles with uniformly distributed rotation axes, Lorentzian frequency distribution with mean frequency $\omega_0=0$ (left and right handed rotations) and width $\gamma$ as a function of the noise strength $D$ (both in units of the coupling strength). The dashed vertical lines mark the critical noise strengths according to our linear stability analysis of the incoherent state, Eqs.~\eqref{Eq:Kcrit},\eqref{Eq:hint}. The solid red line on top of the simulations for $\gamma/K=0$ is the mean field amplitude \eqref{Eq:selfcons} for the globally coupled Vicsek model (von Mises-Fisher distribution).}
	\label{Fig:Rho_vs_DK}
\end{figure}

\subsection{Synchronization/alignment transition}
\label{sec:SynTr}

The simplest case allowing for a full analytic treatment is the one
without oscillations, i.e. $g(\omega)=\delta(\omega)$.  Then the two processes determining the dynamics of vectors $\uvec{\sigma}$ are
diffusion under the influence noise, and alignment  to the mean field $\bvec{\rho}$:
\begin{equation}\label{Eq:SphereDiff}
	\dot{\uvec{\sigma}}_i = \left(K(\uvec{\sigma}_i\times\bvec{\rho})+\bvec{\xi}\right)\times\uvec{\sigma}_i ~.
\end{equation}
The stationary solution of the Fokker-Planck equation \eqref{Eq:FP} can be found analytically. It is current-free, which amounts to a detailed balance condition in \eqref{Eq:FP}
\begin{equation}
	f\bvec{a} = f\cdot K(\bvec{\rho}-(\bvec{\rho}\cdot\uvec{\sigma})\uvec{\sigma})=D\bvec{\nabla}_sf.
\label{Eq:stat_1}
\end{equation}
Without loss of generality, we set $\bvec{\rho}=\rho \uvec{z}$ and multiply both sides of Eq.~(\ref{Eq:stat_1}) by $\uvec{z}$. The resulting one dimensional differential equation for the rotational symmetric density
has the Boltzmann-type von Mises-Fisher distribution as a solution
\begin{equation}\label{Eq:VonMisesFischer}
	f(\theta,\phi)=f(\theta)=\frac{K\rho}{4\pi D \sinh\left(\frac{K\rho}{D}\right)} \exp\left(\frac{K\rho}{D}\cos\theta\right).
\end{equation}
Here $\theta,\phi$ are polar angles defining the direction of the vector $\uvec{\sigma}$.
For this density Eqs.~\eqref{Eq:OrderParmDef} and \eqref{Eq:FamilyOrder}  give the self-consistency condition
\begin{equation}\label{Eq:selfcons}
	\left|\bvec{\rho}\right| = \rho=\coth\left(\frac{K\rho}{D}\right)-\frac{D}{K\rho}.
\end{equation}
Its solution can be represented in a parametric form.
Denoting $x=\frac{K\rho}{D}$, we obtain both the order parameter $\rho$
and the essential parameter of coupling to noise ratio $K/D$ as functions of $x$:
 $\rho=\coth x -1/x$ and $K/D=x^2/(x\coth x - 1)$.
 The auxiliary parameter $0<x<\infty$ varies
 between the transition point at $x\to 0$, where $\rho\to 0$,
 and the noise-free limit $x\to\infty$, where $\rho\to 1$ (complete alignment). 
 These analytic expressions are in agreement with direct simulations of  Eq.~\eqref{Eq:SphereDiff}, as depicted in Fig.~\ref{Fig:Rho_vs_DK} (case $\gamma/K=0$). Expanding Eq.~(\ref{Eq:selfcons}) to the third order in $\rho$ 
we obtain close to the transition point
\begin{equation}
	\rho\approx\frac{K\rho}{3D}-\frac{K^3\rho^3}{45D^3}, \quad\textrm{or}\quad \rho\approx\sqrt{\frac{15D^2(K-3D)}{K^3}},
\end{equation}
i.e. the globally coupled Vicsek model has a continuous transition at $K=3D$ with  critical exponent $1/2$.

\subsection{A model for von Mises-fisher distribution of rotation axes}
Above in Section~\ref{sec:SynTr} we considered a simple situation without rotation biases. Below we perform a more general analysis that includes distributions of the frequencies  $g(\omega)$
and of the rotation axes $G(\uvec{\omega})$. The latter
is a distribution on a sphere, and it is natural to assume it belongs to
the von Mises-Fisher family of distributions (because this family spans a range from the uniform to a very narrow distribution). As it follows from the analysis
above, a von Mises-Fisher distribution naturally appears as a
stationary distribution for the Langevin process \eqref{Eq:SphereDiff}.
Therefore, below we use the model where rotation axes
$\uvec{\omega}_i$ are not constants, but evolve slowly like
in  \eqref{Eq:SphereDiff}:
\begin{equation}\label{Eq:FreqDiff}
	\dot{\uvec{\omega}}_i = \left(\kappa (\uvec{\omega}_i\times\langle \uvec{\omega}\rangle)+\bvec{\zeta}\right)\times\uvec{\omega}_i ~.
\end{equation}
where $\bvec{\zeta}$ with $\left\langle (\bvec{\zeta}_i)_n(t)(\bvec{\zeta}_j)_m(t')\right\rangle = 2d\delta_{ij}\delta_{mn}\delta(t-t')$ is Gaussian white noise. 
If the coupling $\kappa$ and the noise intensity $d$ are small,
the evolution of the distribution  $G(\uvec{\omega},t)$ according
to \eqref{Eq:FreqDiff} is slow. Furthermore, as will be illustrated below, during this evolution   $G(\uvec{\omega},t)$ is a slowly evolving von Mises-Fisher distribution. This is confirmed in
Fig.~\ref{Fig:Examples} below by monitoring the ensemble moments
$\langle\hat\omega_z\rangle$, $\langle \hat{\omega}_z^2\rangle$, $\langle \hat{\omega}_x^2\rangle$ and $\langle \hat{\omega}_y^2\rangle$. According to \eqref{Eq:selfcons} for a von Mises-Fisher distribution \eqref{Eq:VonMisesFischer} of rotation axes
\begin{equation}\label{Eq:FreqVonMisesFischer}
	f_\omega(\theta,\phi)=\frac{\kappa\langle \hat{\omega}_z\rangle}{4\pi d \sinh\left(\frac{\kappa\langle \hat{\omega}_z\rangle}{d}\right)} \exp\left(\frac{\kappa\langle \hat{\omega}_z\rangle}{d}\cos\theta\right)\;,
\end{equation}
which for $d<3\kappa$ has the second moments
\begin{eqnarray}
    \langle\hat\omega_z^2\rangle &=& 1- 2d/\kappa\;, \\
    \langle\hat\omega_x^2\rangle &=& \langle\hat\omega_y^2\rangle = d/\kappa\;,
\end{eqnarray}
the deviation
\begin{equation}\label{Eq:vMF_Distance}
    \Delta = \langle \hat\omega_z\rangle + \frac{\langle\hat\omega_x^2\rangle}{\langle\hat\omega_z\rangle} - \coth\left(\frac{\langle\hat\omega_z\rangle}{\langle\hat\omega_x^2\rangle}\right)
\end{equation}
must be zero. We check numerically, that in our simulations 
this is not only valid in the final
stationary state, but also during the transient. This allows us to
study the synchronization transition in an adiabatically evolving von Mises-Fisher distribution of the rotation axes.

\section{Linear stability analysis of the incoherent state}
\label{sec:linear stablity}
In the following we analyse the stability of the incoherent state where the vectors $\uvec{\sigma}$ (or velocities $\uvec{v}$) are distributed uniformly in all directions and $|\bvec{\rho}| = \rho = 0$. 
Following the non-trivial derivation in \cite{chandra2019continuous}, the Fokker-Planck equation (\ref{Eq:FP}) can be rewritten as
\begin{equation}
	\frac{\partial f}{\partial t}+K(\bvec{\nabla}_s f-2f\uvec{\sigma})\cdot\bvec{\rho}
	+(\omega\uvec{\omega}\times\uvec{\sigma})\cdot\bvec{\nabla}_sf=D\bvec{\nabla}^2_sf.
\label{Eq:FP_re}
\end{equation}
We consider a small perturbation on top of
the  uniform incoherent distribution $f_0=(4\pi)^{-1}$.
Substituting the ansatz $f=f_0+\eta(\uvec{\sigma},\uvec{\omega},\omega)e^{st}$ for a small perturbation into Eq.~(\ref{Eq:FP_re}) and assuming without loss of generality $\uvec{\omega}=\uvec{z}$ (this allows us to express the eigenmode in terms of angles $\theta,\phi$), we obtain to the linear order in $\rho$ and $\eta$ the equation
\begin{eqnarray} \label{Eq:character}
	2K(\bvec{\rho}\cdot\uvec{\sigma})(4\pi)^{-1} &=& \omega\frac{\partial}{\partial\phi}\eta(\theta,\phi,\omega) \\
	&+& s\eta(\theta,\phi,\omega)-D\bvec{\nabla}^2_s
	\eta(\theta,\phi,\omega)\;.\nonumber
\end{eqnarray}

In order to solve Eq.~(\ref{Eq:character}), we express $\uvec{\sigma}$ and $\eta(\theta,\phi,\omega)$ in terms of bi-orthonormal spherical harmonics $Y_l^m(\theta,\phi)$ as
\begin{equation}\label{Eq:sigmaYlm}
	\uvec{\sigma} = \sqrt{\frac{2\pi}{3}}\left(\begin{array}{c}
												Y_{1}^{-1} - Y_{1}^{1} \\
												iY_{1}^{-1} + iY_{1}^{1}\\
												\sqrt{2}Y^0_1
										  \end{array}\right)
\end{equation}
and as
\begin{equation}
	\eta(\theta,\phi,\omega)=\sum\limits_{l=0}^{\infty}
	\sum\limits_{m=-l}^{l}b^{m}_{l}(\omega) Y^{m}_{l}.
\label{Eq:ansatz}
\end{equation}
On the surface of the sphere the action of the diffusion term reduces to
$\bvec{\nabla}^2_s Y^{m}_{l}(\theta,\phi)=-l(l+1)Y^{m}_{l}(\theta,\phi)$.
Substituting this expansion into
Eq.~(\ref{Eq:character}), we obtain
a linear system of equations for the coefficients $b^{m}_{l}$
\begin{equation}
	\begin{aligned}
		&\frac{1}{\sqrt{6\pi}}K\left(\sqrt{2}\rho_z Y_1^0 + (\rho_x+i\rho_y)Y_1^{-1} -(\rho_x-i\rho_y)Y_1^{1} \right) \\
		&=\sum\limits_{l=0}^{\infty}\sum\limits_{m=-l}^{l}b^{m}_{l}[s+im\omega+Dl(l+1)]Y^{m}_{l}(\theta,\phi)\;,
	\end{aligned}
\label{Eq:harmexpan}
\end{equation}
which can be solved using the orthonormality of the spherical harmonics. Since the l.h.s. depends on $Y_l^m$ with $l=1$ only, the components $b_l^m$ with $l>1$ decay exponentially at rates $Dl(l+1)$. For $l=1$ the coefficients $b_1^0$, $b_1^{-1}$ and $b_1^1$ are calculated explicitly, resulting in the general form
\begin{equation}
	\begin{aligned}
		\eta(\theta,\phi,\omega)&=\sqrt{\frac{1}{3\pi}}\frac{K\rho_z}{s+2D}Y^0_1+\sqrt{\frac{1}{6\pi}}\frac{K(\rho_x+i\rho_y)}{s			+2D-i\omega}Y^{-1}_1\\&-\sqrt{\frac{1}{6\pi}}\frac{K(\rho_x-i\rho_y)}{s+2D+i\omega}Y^{1}_1
	\end{aligned}
\label{Eq:solu_pt}
\end{equation}
of a potentially unstable mode. Integrating $\uvec{\sigma}\eta$ over the surface of the sphere (Eqs.~\eqref{Eq:FamilyOrder},\eqref{Eq:sigmaYlm} and \eqref{Eq:solu_pt}), using again 
the orthonormality of the spherical harmonics, we obtain the frequency dependent mean fields (moments of the linear perturbation) 
to linear order
\begin{equation}
	\bvec{r} = \int_{S^1} \uvec{\sigma}\eta(\theta,\phi,\omega)dA(\uvec{\sigma})
		=\frac{2K}{3}\begin{pmatrix} \frac{\rho_x\lambda-\rho_y\omega}{\lambda^2+\omega^2} \\ \frac{\rho_y\lambda+\rho_x\omega}{\lambda^2+\omega^2}\\\frac{\rho_z}{\lambda},\end{pmatrix}\;,
\label{Eq:R_omegarho1}
\end{equation}
where $\lambda=s+2D$.
According to the convention above, $\uvec{\omega}$ is directed along $\uvec{z}$ while the direction of $\bvec{\rho}$ is arbitrary.
However, expression \eqref{Eq:R_omegarho1} can be rewritten in a covariant form, allowing for arbitrary directions of $\bvec{\rho}$ and $\uvec{\omega}$
\begin{equation}\label{Eq:R_omegarho}
	\bvec{r} = \frac{2K}{3} \left[\frac{\lambda\bvec{\rho}}{\lambda^2+\omega^2} + \frac{\omega \uvec{\omega}\times\bvec{\rho}}{\lambda^2+\omega^2}+ \left(\frac{1}{\lambda}-\frac{\lambda}{\lambda^2+\omega^2}\right)\uvec{\omega}\left(\uvec{\omega}\cdot\bvec{\rho}\right)\right].
\end{equation}

To express the resulting dispersion relation equation, it is convenient
to introduce the following notations:\\
(i) we introduce the averages over
the distribution of the frequencies as
\begin{equation}
	h_1 = \int\limits_{-\infty}^\infty \frac{\lambda}{\lambda^2+\omega^2}g(\omega)d\omega,
	\quad h_2=\int\limits_{-\infty}^\infty \frac{\omega}{\lambda^2+\omega^2}g(\omega)d\omega,
\label{Eq:h_1_h_2_h_3}
\end{equation}
and $h_3=\frac{1}{\lambda}-h_1$;\\
(ii) we introduce two matrices, characterizing the distribution
of the rotation axes: the antisymmetric matrix
of the first moments $\mathrm{\Xi}$ as
\begin{equation}
	\mathrm{\Xi}=\int_{S^1} \left(\begin{array}{ccc}
									0& -\hat{\omega}_z& \hat{\omega}_y\\
									\hat{\omega}_z& 0& -\hat{\omega}_x\\
									-\hat{\omega}_y& \hat{\omega}_x& 0
                            \end{array}\right) G(\uvec{\omega})dA(\uvec{\omega})
\label{Eq:mXi}
\end{equation}
and the covariance matrix $\mathrm{W}$ as
\begin{equation}
	\mathrm{W} =\int_{S^1} \left(\begin{array}{ccc}
									\hat{\omega}^2_x& \hat{\omega}_x\hat{\omega}_y& \hat{\omega}_x\hat{\omega}_z\\
									\hat{\omega}_x\hat{\omega}_y& \hat{\omega}^2_y& \hat{\omega}_y\hat{\omega}_z\\
									\hat{\omega}_x\hat{\omega}_z& \hat{\omega}_y\hat{\omega}_z& \hat{\omega}^2_z
                                 \end{array}\right) G(\uvec{\omega})dA(\uvec{\omega})\;.
\label{Eq:mW}
\end{equation}

With these notations
we can express $\bvec{\rho}$ from (\ref{Eq:OrderParmDef}) and (\ref{Eq:R_omegarho}) self-consistently in a compact form
\begin{equation}\label{Eq:EigenValue}
	\bvec{\rho} = \frac{2K}{3}\left[h_1 \mathbb{1} + h_2 \mathrm{\Xi} + h_3 \mathrm{W}\right] \bvec{\rho}.
\end{equation}
The real part of the exponent $s=\lambda-2D$ for any mode $\bvec{\rho}$ matching this eigenvalue equation gives the growth rate of that mode. Equation (\ref{Eq:EigenValue}) has a non trivial solution $\bvec{\rho}$ if the dispersion relation
\begin{equation}\label{Eq:DispersionRelation}
	\textrm{det}\left[\frac{2K}{3}\left(h_1 \mathbb{1} + h_2 \mathrm{\Xi} + h_3 \mathrm{W}\right)-\mathbb{1}\right] = 0
\end{equation}
holds. This is the main result of our paper and we will discuss consequences and examples in the following sections. But first we would like to examine general properties of Eq. \eqref{Eq:DispersionRelation}.
Because both the real and the imaginary part of the 
determinant~\eqref{Eq:DispersionRelation} must be zero at 
criticality where $s=i\Omega$ and other system parameters are fixed, this 
occurs at a discrete set of points $(K_l,\Omega_l)$ (see an example in Fig. \ref{Fig:DetSolve} below). 
At the smallest coupling strength $K_{cr} = \textrm{min}_l K_l$, the incoherent state loses stability and a nonzero mean field with 
frequency $\Omega_{cr}$ emerges. For any critical mode with $(K_l,\Omega_l)$ the mode with $(K_l,-\Omega_l)$ is also critical. Moreover, there is always at least one non-oscillating
solution $(K_l,\Omega_l=0)$ since the determinant is a cubic polynomial in $K$ with real coefficients when $\Omega=0$. A nonzero frequency $\Omega_{cr}$ at the bifurcation indicates the formation 
of a rotating velocity mean field in the swarming model where the variables are interpreted as velocities $\uvec{\sigma}=\uvec{v}$.
This means that the population will demonstrate
 coherent oscillations. In contradistinction, if the critical mode
 has zero frequency $\Omega_{cr}=0$, a transition to a regime with
 a stationary non-zero mean velocity occurs. 
 This corresponds to a directed motion of the swarm's  center of mass.

The dependence of the real and imaginary parts of the matrix determinant in \eqref{Eq:DispersionRelation} on the system parameters can be arbitrary complicated (see Fig.\ref{Fig:DetSolve}c). Changing the system parameters, pairs of points $(K_l,\Omega_l)$ can emerge or annihilate and the sequence of critical coupling strengths for these unstable modes, and thus the type of the emerging collective motion can change.

\section{Synchronization in the presence of a uniformly distributed rotation bias}
\label{subsec:Disordered case}
In the presence of individual, quasi-static rotation axes, the model described by Eq.~(\ref{Eq:SphereForce}) and (\ref{Eq:Torque}) is a noisy version of the recently proposed three-dimensional generalization of the Kuramoto model \cite{chandra2019continuous}. Indeed, in two dimensions the connection between the Vicsek model and the Kuramoto model has been made explicit \cite{chepizhko2010relation,degond2014hydrodynamics}. The three-dimensional Kuramoto model without noise was discussed as a swarming model in \cite{chandra2019continuous}. Strikingly and in stark contrast to the classical Kuramoto model, despite heterogeneous frequency amplitudes and rotation directions, which were described in \cite{chandra2019continuous} as imperfections that make individuals deviate from ideally straight lines, global coupling leads to a finite translational collective motion for arbitrary small coupling strength, when all oscillations cease as the velocities settle at well-defined fixed points. Frequency heterogeneity is not sufficient to prevent velocity alignment. 

On the other hand, random perturbations of the torque in form of Gaussian white noise stabilize the incoherent state, much as in the classical Kuramoto model, and a transition to collective motion occurs at finite coupling strength. In Fig.~\ref{Fig:Rho_vs_DK} we show the mean velocity as a function of the relative noise strength $D/K$ for an isotropic distribution of rotation axes $G(\uvec{\omega})=1/(4\pi)$ and Lorentzian frequency distributions $g(\omega)=\frac{\gamma}{\pi}(\omega^2+\gamma^2)^{-1}$ with mean frequency zero and width $\gamma$ characterizing the frequency heterogeneity. Depending on the ratio $\gamma/K$ a stationary mean field bifurcates continuously from $\rho=0$ at a critical value of $(D/K)_{cr}$, which the linear stability analysis in Section \ref{Sec:ge} predicts. 

The branch of partially
synchronized states stretches from this bifurcation point on the horizontal axis  (where $\rho=0$) 
to a point on the vertical axis (at $D=0$) in the noise free limit, discussed in \cite{chandra2019continuous}. The existence of a critical ratio $(D/K)_{cr}$ for the transition from incoherence to coherence means that the noise free limit $D\to 0$ is singular as the critical coupling strength also goes to zero. In Fig.~\ref{Fig:Rho_vs_DK} we show four examples with different frequency heterogeneities $\gamma/K=0, 0.1, 1$ and $10$. The rotation-free case $\gamma/K=0$ corresponds to the globally coupled Vicsek model for which the bifurcation curve is known parametrically (Eq. \eqref{Eq:selfcons}, solid red line). When the frequencies are very heterogeneous, e.g. $\gamma/K=10$, the incoherent state, where the mean velocity is zero, is stable for even lower ratios of $D/K$. For $D=0$ the mean velocity in the limit $\gamma/K\to\infty$ is $\rho=0.5$ corresponding to the limit $K\to 0$ as predicted in \cite{chandra2019continuous}.

\section{Axial-symmetric distribution of rotating axes}
\label{subsec:Symmetric frequency}

In this section we go beyond the simplest 
setup of Section~\ref{subsec:Disordered case} and discuss a nontrivial
situation where there is a preferable direction of rotation axes $\uvec{\omega}$.
\begin{figure}[ht!]
\setlength{\unitlength}{1cm}
\begin{picture}(3.8,3.8)
\put(-0.5,0){\includegraphics[height=3.8cm]{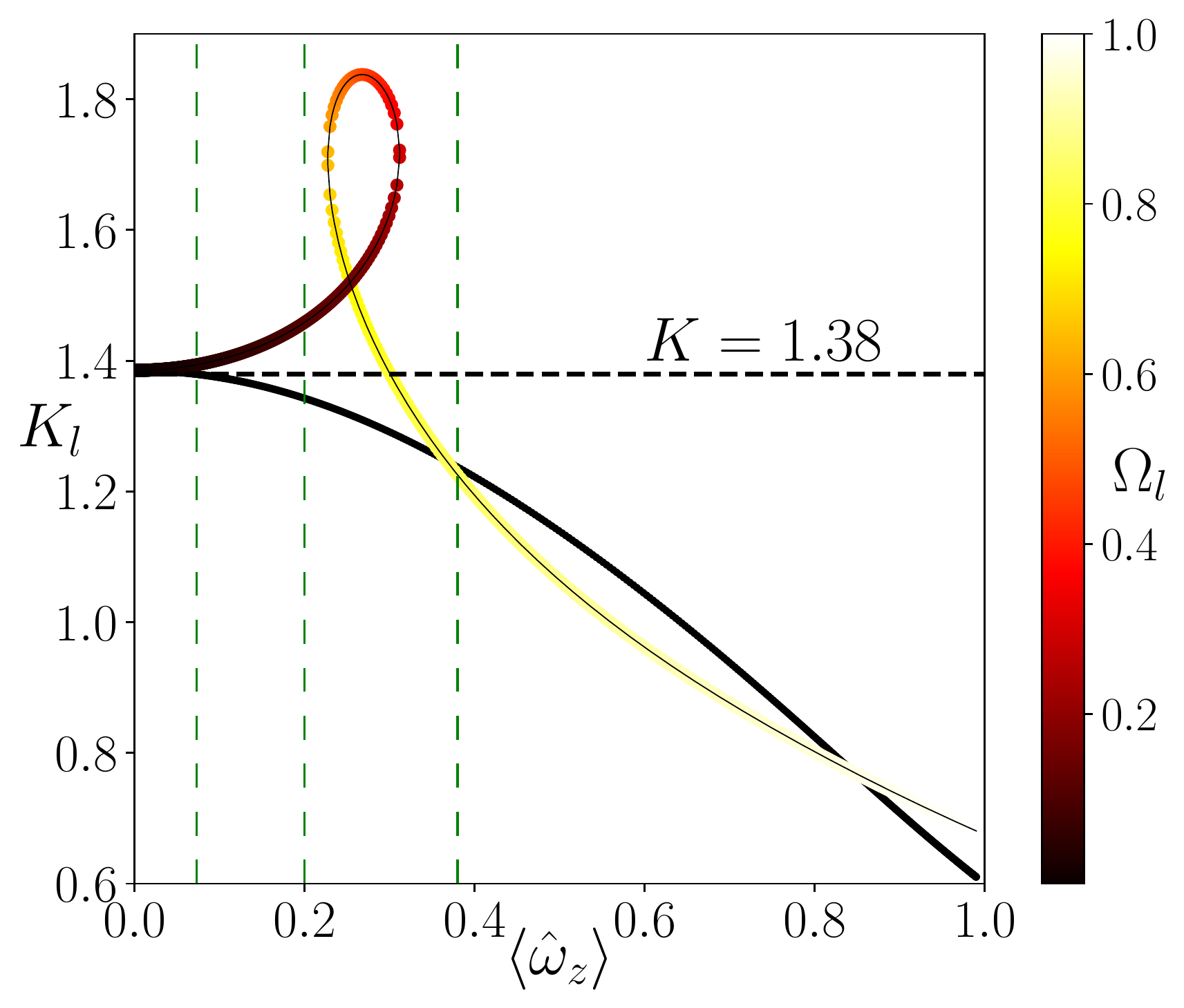}}
\put(-0.8,3.4){\bf (a)}
\end{picture}
\begin{picture}(3.8,3.8)
\put(0.3,0){\includegraphics[width=3.8cm]{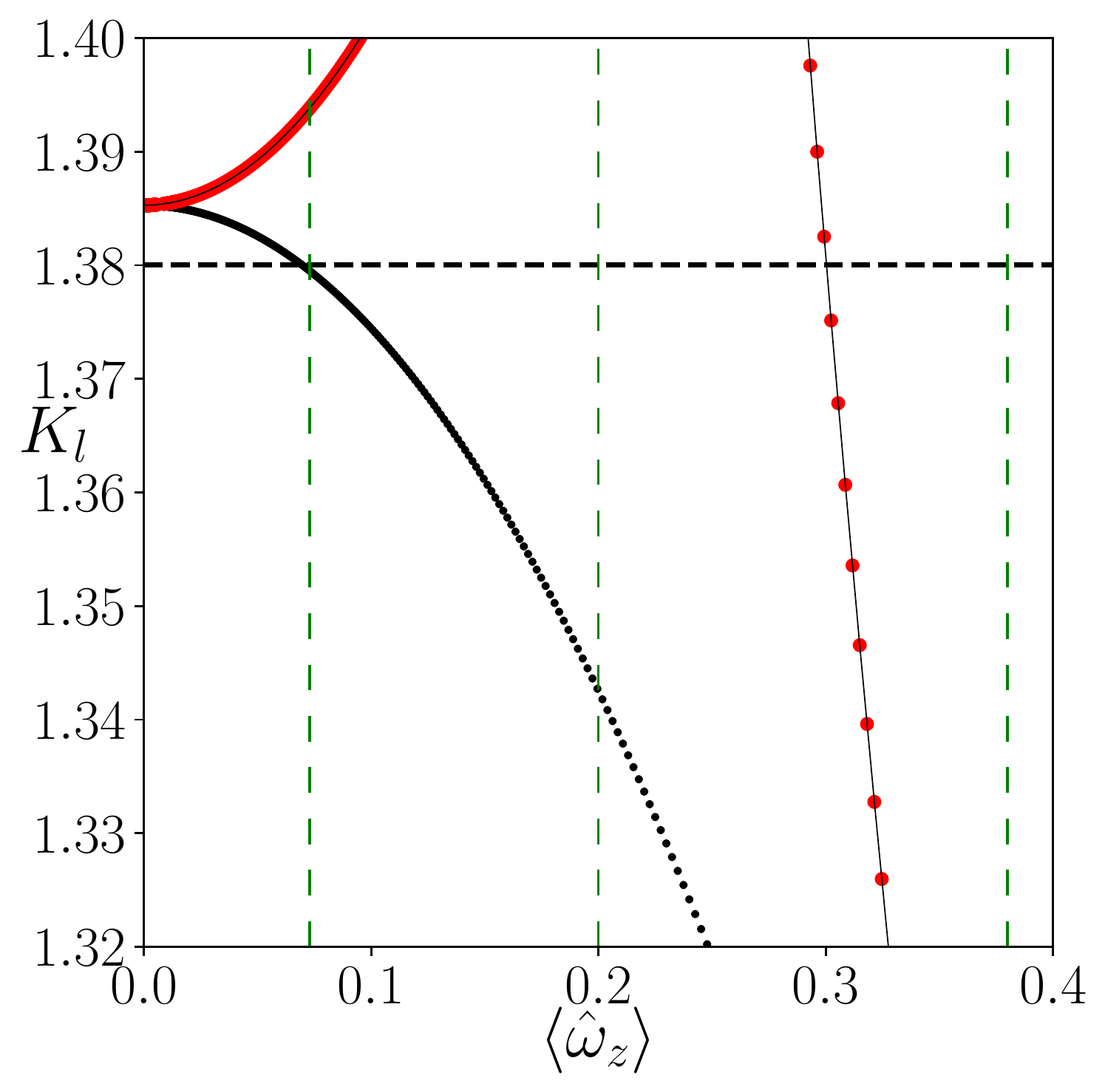}}
\put(-0.1,3.4){\bf (b)}
\end{picture}
\begin{picture}(7.6,4.2)
\put(-0.5,0){\includegraphics[width=7.6cm]{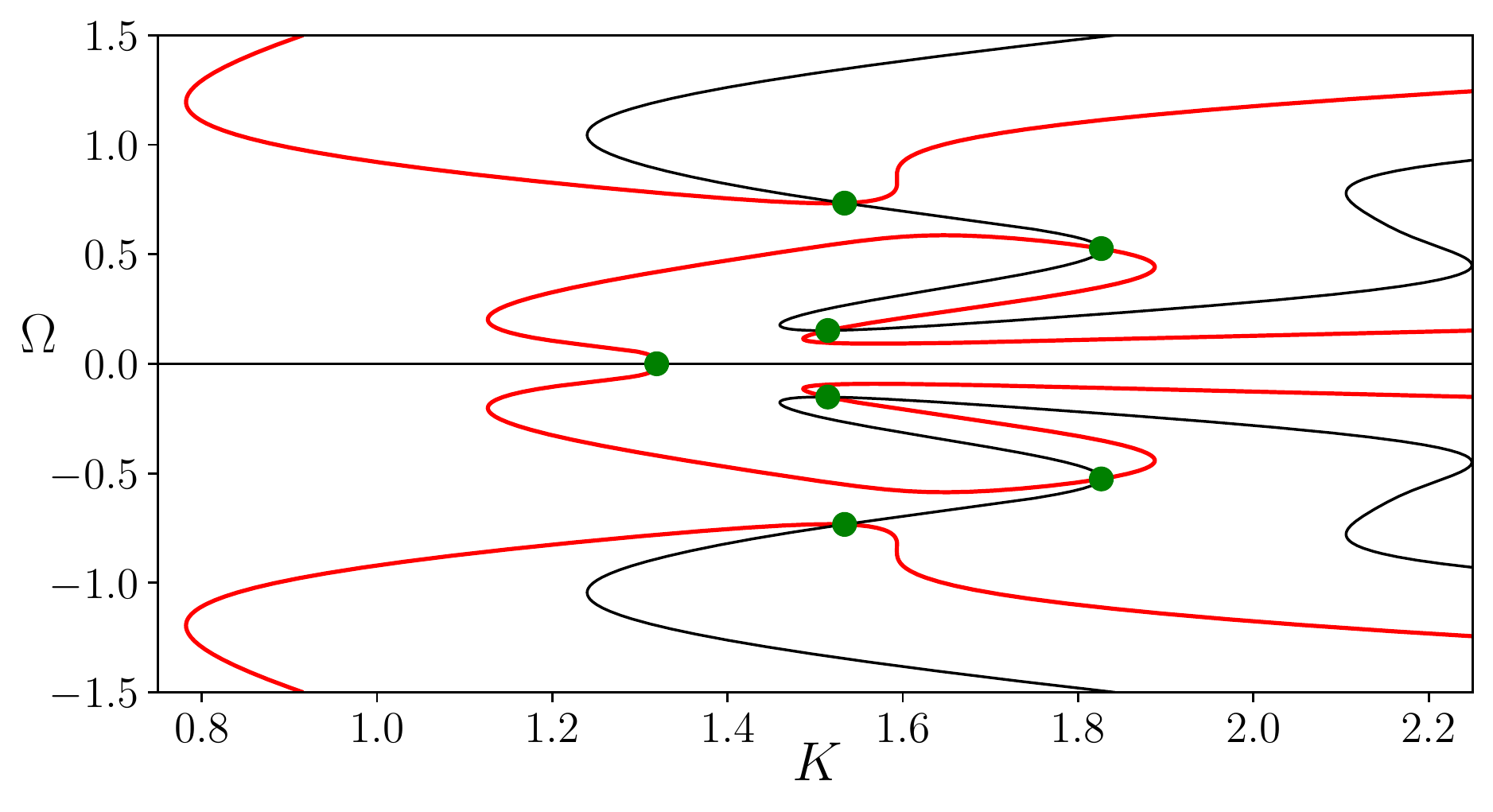}}
\put(-0.5,4){\bf (c)}
\end{picture}
\caption{Solutions $(K_l,\Omega_l)$ of dispersion relation \eqref{Eq:DispersionRelation} in (a-b) as a function of the rotation axes mean field amplitude $\langle \hat\omega_z\rangle$ (von Mises-Fisher distribution). Panel (c) shows the roots of the dispersion relation on the $(K,\Omega)$ plane for $\langle \hat\omega_z\rangle=0.25$. The color shade on the two branches in (a) denotes the frequency $\Omega_l$ of the corresponding unstable mode. We see one branch having frequency zero (bold black line/dots), corresponding to a stationary directed mean velocity, and another branch of oscillatory instabilities (thin black line and colored circles, correspondingly). Depending on $K$ one of these two types of instabilities occurs first when $\langle \hat\omega_z\rangle$ increases. The dashed horizontal line marks the coupling $K=1.38$ in the examples of Fig.\ref{Fig:Examples}. The dashed vertical lines mark the values $\langle \hat{\omega}_z\rangle=0.08$, $0.2$ and $0.38$ corresponding to the horizontal lines in Fig.~\ref{Fig:Examples}a. The first instability at $\langle \hat{\omega}_z\rangle=0.08$ corresponds to a directed motion, the second, oscillatory 
unstable mode appears at $\langle \hat{\omega}_z\rangle=0.31$. This can be seen in the magnified plot in (b).   At $\langle \hat{\omega}_z\rangle=0.25$ we show in 
panel (c) critical coupling values for the unstable modes and corresponding frequencies (dark green circles). At these points the real part (light red lines) and the imaginary part  (black lines) 
of the dispersion
relation Eq.~\eqref{Eq:DispersionRelation} vanish simultaneously. The three oscillatory modes are part of the same (colored) branch in panel (a).}
	\label{Fig:DetSolve}
\end{figure}
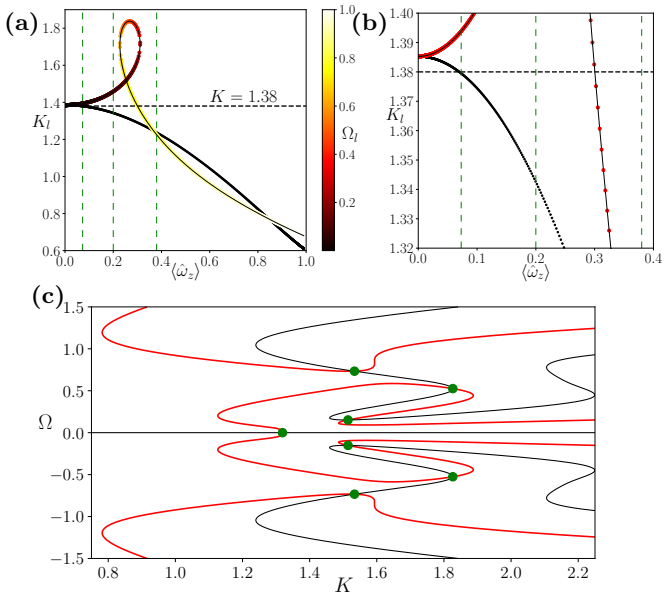

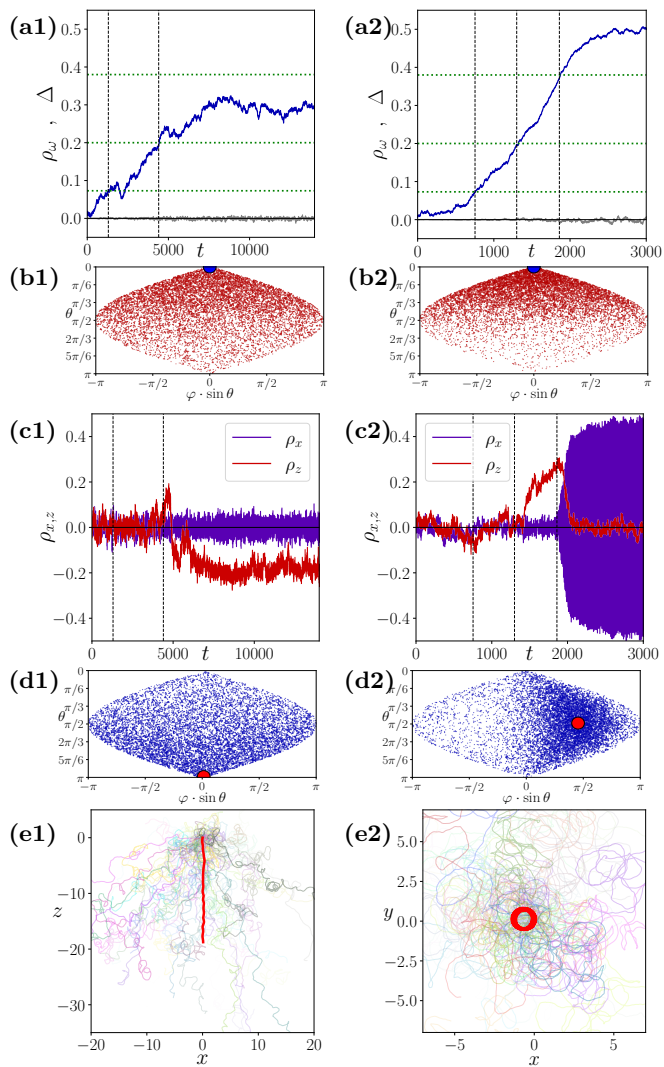
\begin{figure}[ht!]
\setlength{\unitlength}{1cm}
\begin{picture}(3.8,3.5)
\put(-0.5,0){\includegraphics[height=3.5cm]{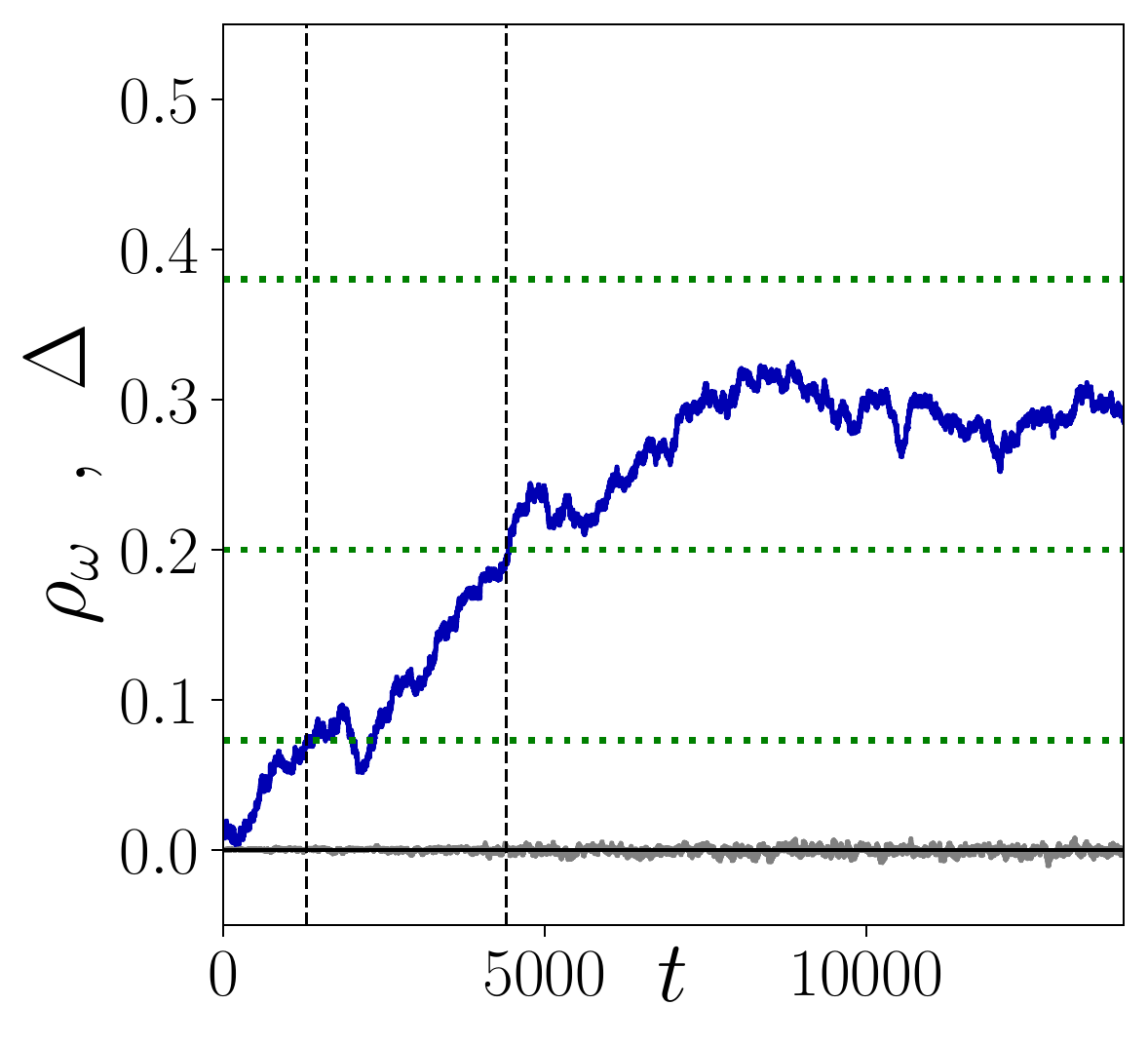}}
\put(-0.8,3.1){\bf (a1)}
\end{picture}
\begin{picture}(3.8,3.5)
\put(0,0.0){\includegraphics[height=3.5cm]{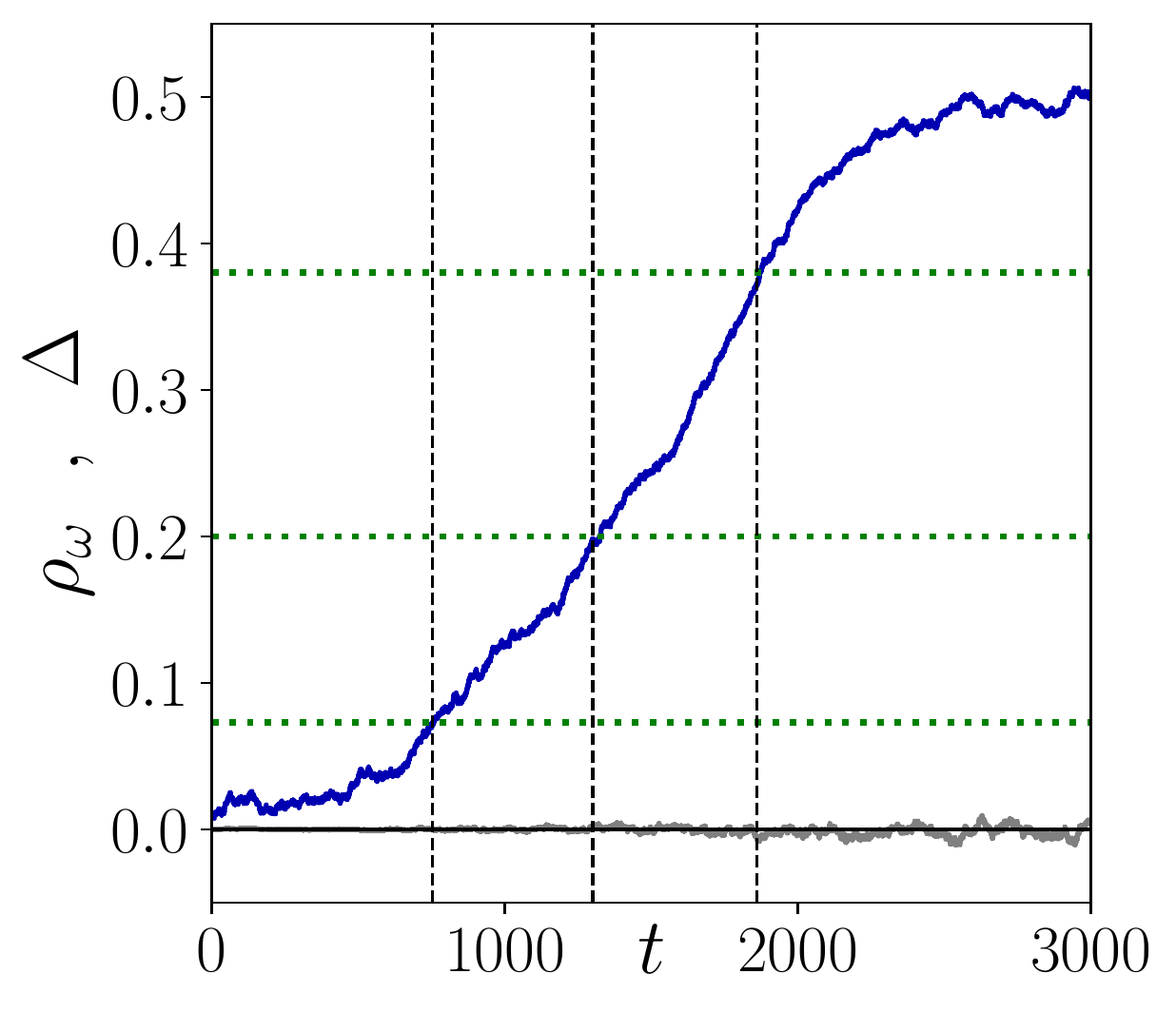}}
\put(-0.25,3.1){\bf (a2)}
\end{picture}
\begin{picture}(3.8,1.8)
\put(-0.1,0){\includegraphics[width=3.6cm, height=1.95cm]{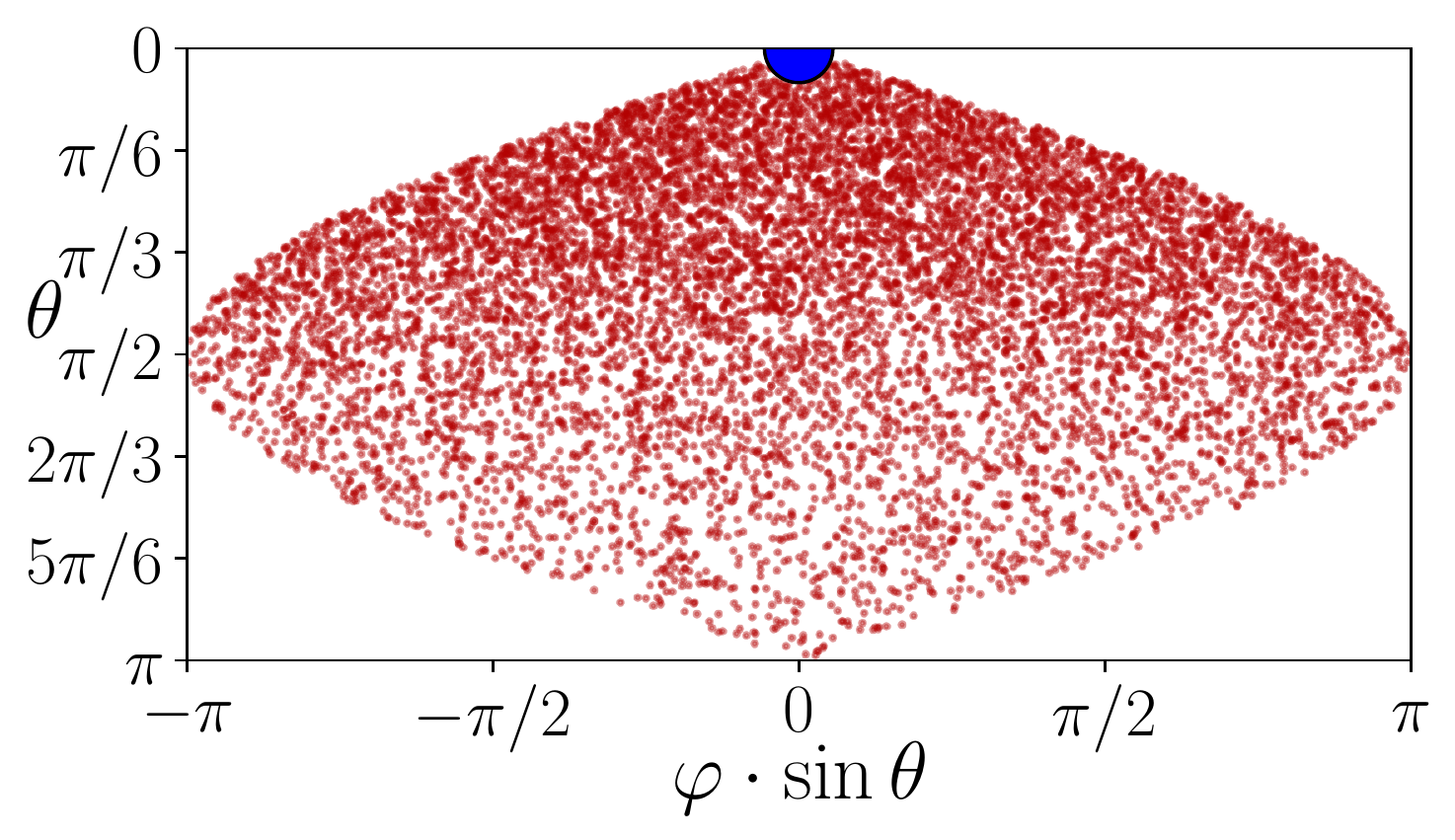}}
\put(-0.8,1.6){\bf (b1)}
\end{picture}
\begin{picture}(3.8,1.8)
\put(0.3,0){\includegraphics[width=3.6cm, height=1.95cm]{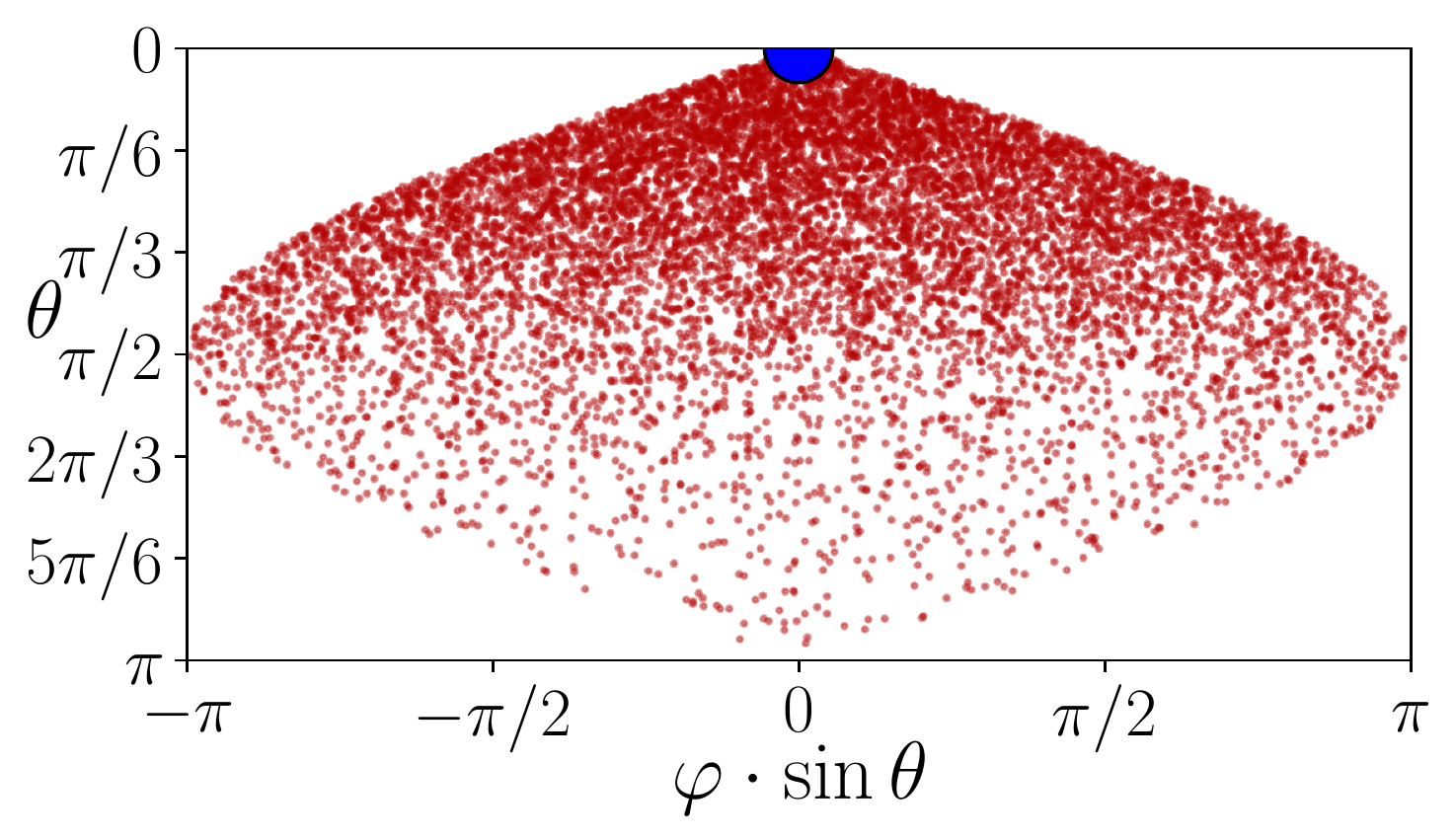}}
\put(-0.2,1.6){\bf (b2)}
\end{picture}
\begin{picture}(3.8,3.5)
\put(-0.5,0){\includegraphics[height=3.5cm]{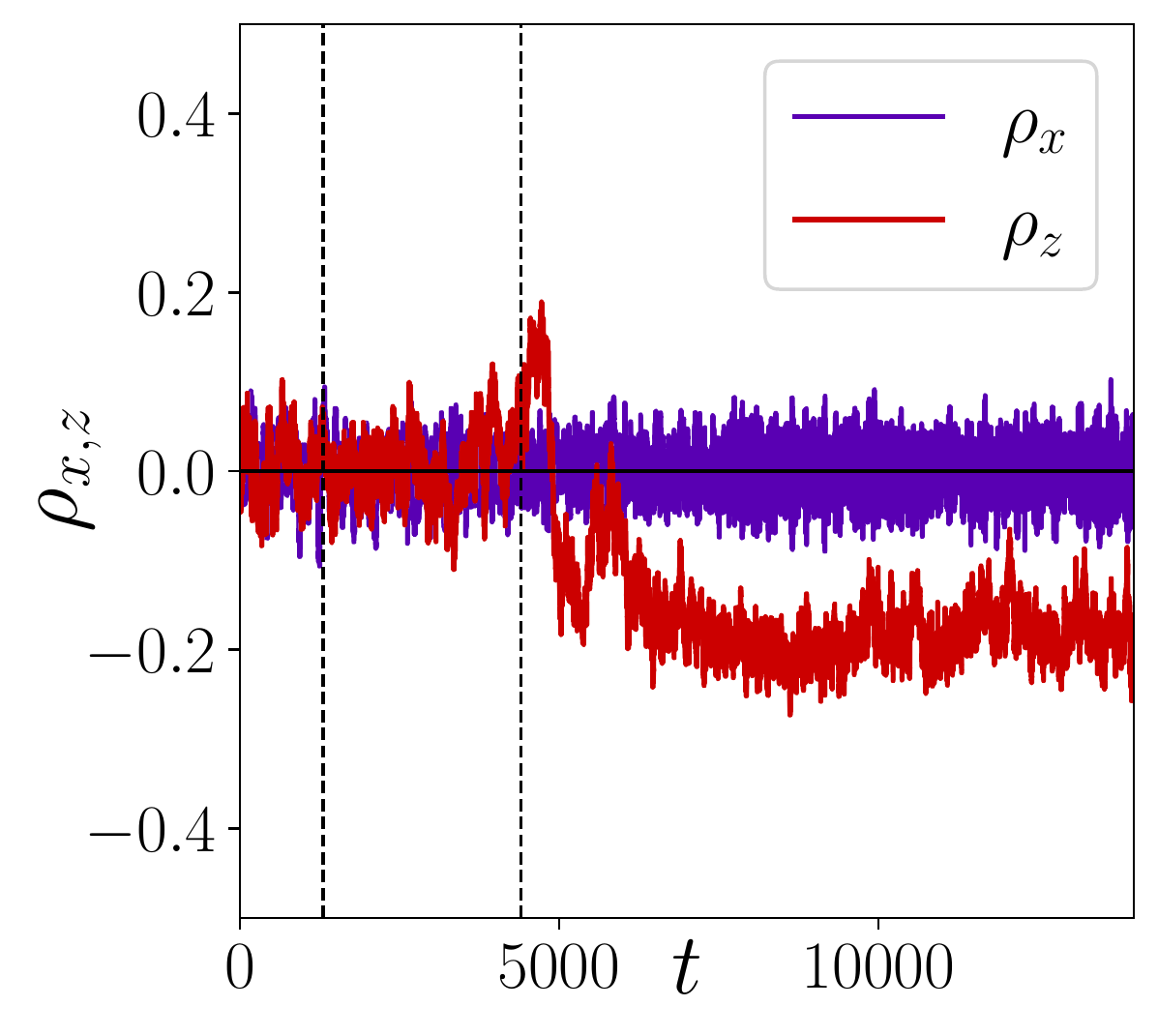}}
\put(-0.8,3.1){\bf (c1)}
\end{picture}
\begin{picture}(3.8,3.5)
\put(-0.1,0){\includegraphics[height=3.5cm]{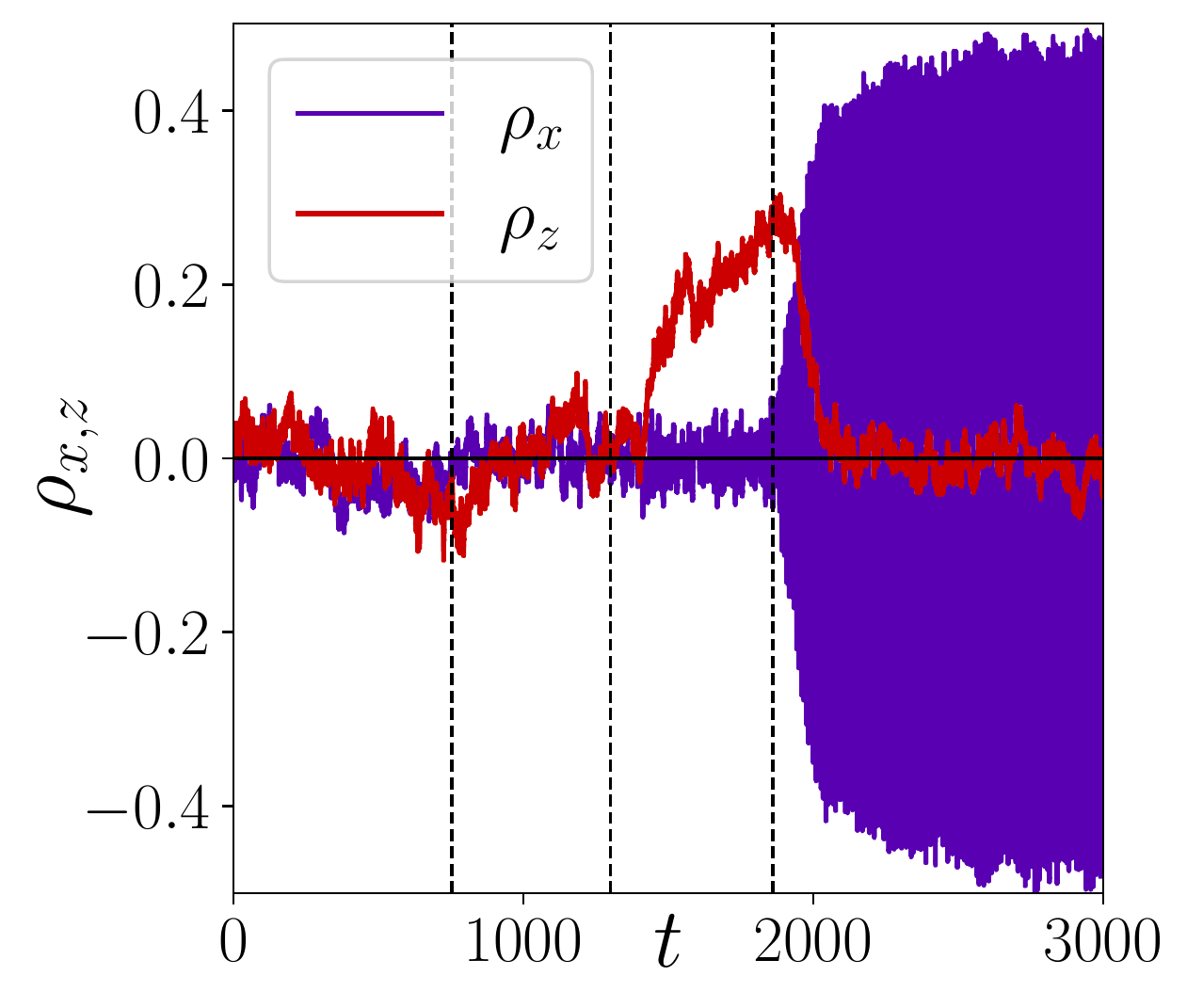}}
\put(-.25,3.1){\bf (c2)}
\end{picture}
\begin{picture}(3.8,1.8)
\put(-0.2,0){\includegraphics[width=3.6cm, height=1.95cm]{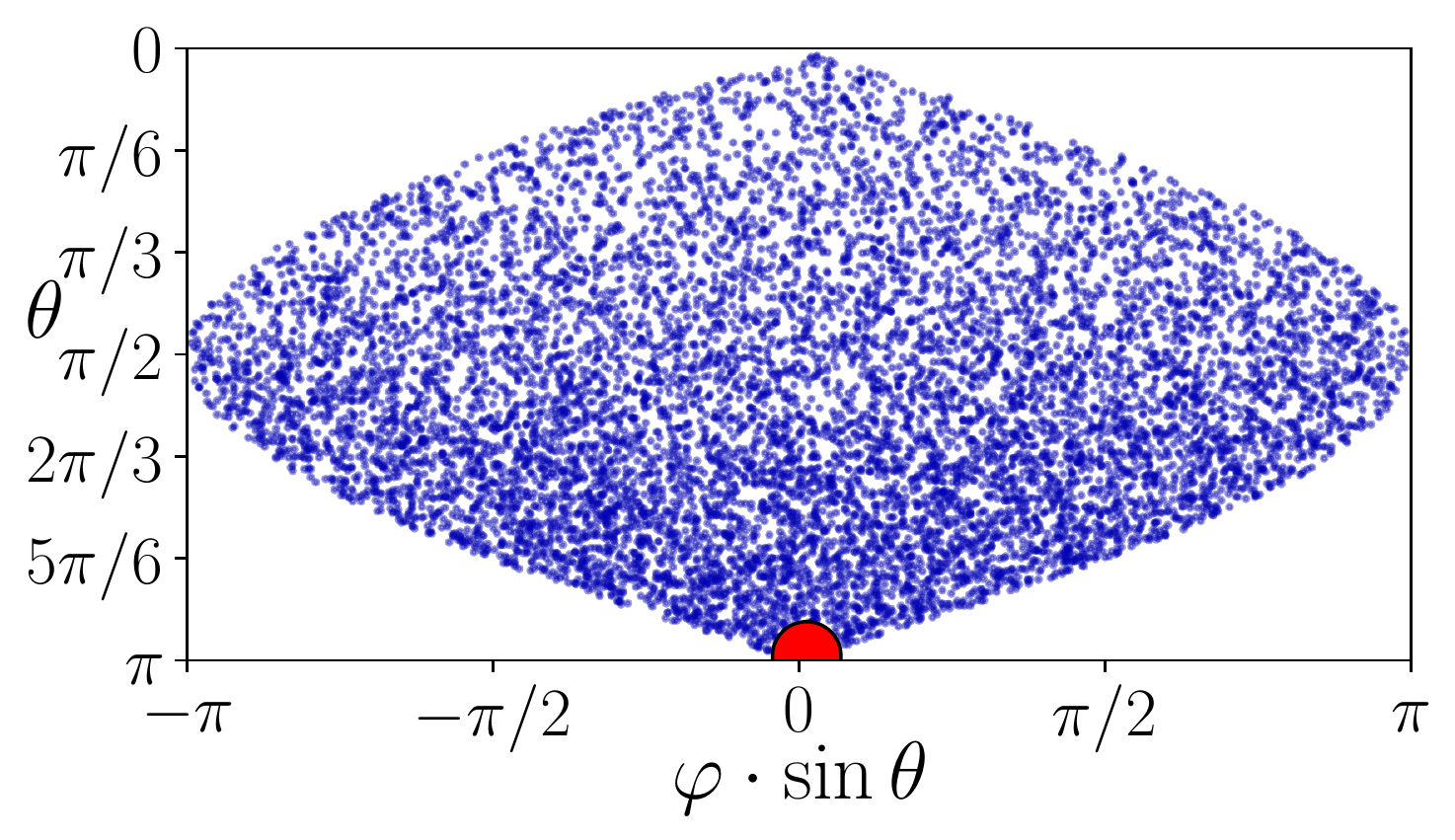}}
\put(-0.8,1.6){\bf (d1)}
\end{picture}
\begin{picture}(3.8,1.8)
\put(0.2,0){\includegraphics[width=3.6cm, height=1.95cm]{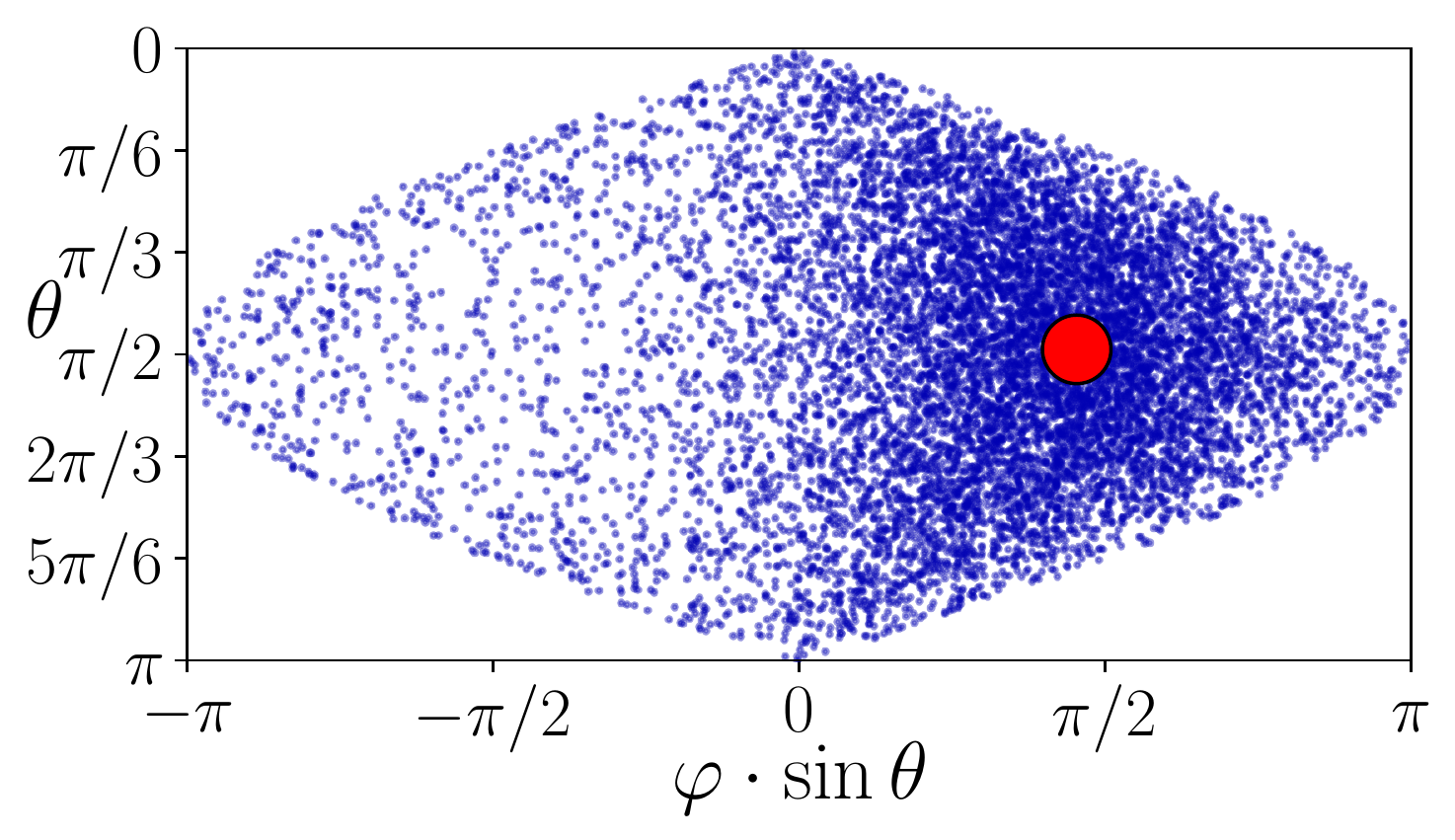}}
\put(-0.25,1.6){\bf (d2)}
\end{picture}
\begin{picture}(3.8,3.5)
\put(-0.3,0){\includegraphics[height=3.6cm]{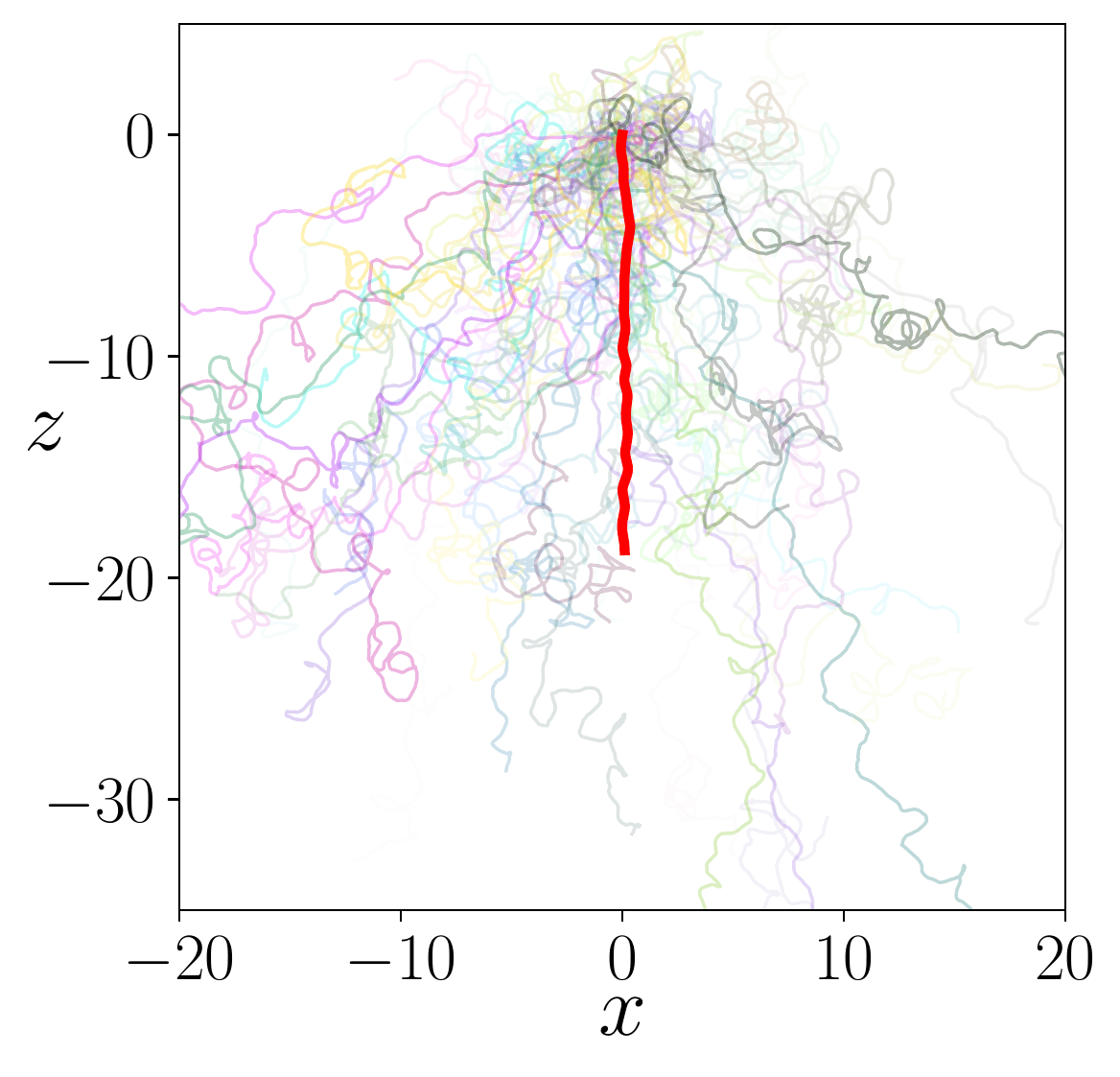}}
\put(-0.8,3.1){\bf (e1)}
\end{picture}
\begin{picture}(3.8,3.5)
\put(0.2,0){\includegraphics[height=3.6cm]{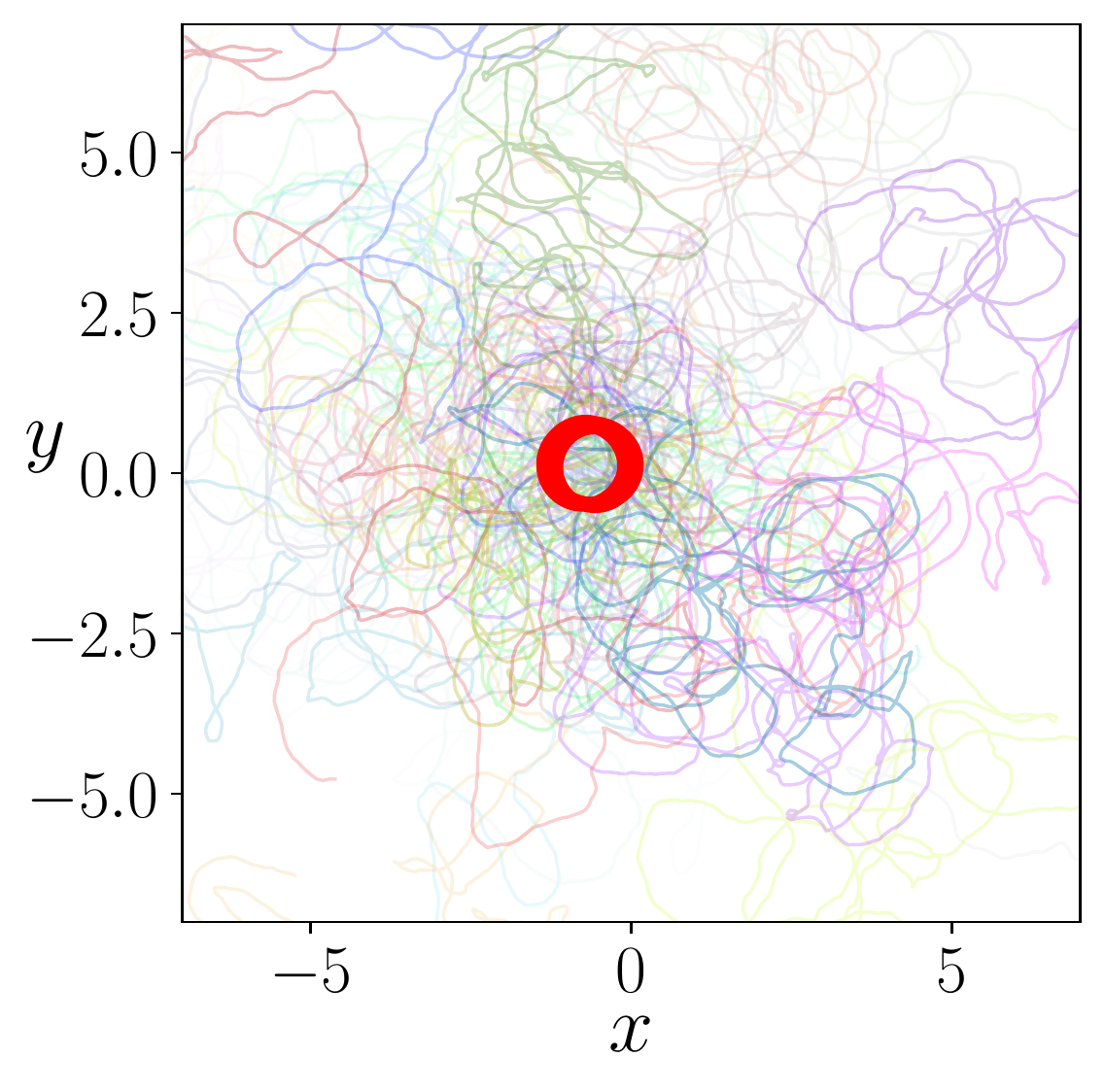}}
\put(-0.25,3.1){\bf (e2)}
\end{picture}
\caption{
Transient and final states of $N=10000$ particle velocities and frequency vectors with $K=1.38$, noise level $D=0.2$, mean rotation frequency $\omega_0=1.0$ and frequency heterogeneity $\gamma=0.05$. The directions of the rotation axes evolve from uniformly random initial conditions under the influence of mean field coupling $\kappa$ and angular diffusion $d=0.005$. In the left column $\kappa=0.0158$ and in the right column $\kappa=0.018$. Panels from top to bottom show: (a) the mean field amplitude $\langle \hat\omega_z\rangle$ of the rotation axes (rising, dark blue curves) and the deviation $\Delta$ from the von Mises-Fisher distribution (Eq.~\ref{Eq:vMF_Distance}, flat, light curves) as functions of time; (b) the final stationary distribution of rotation axes directions as small dots on the unit sphere (sinusoidal projection) and a large circle in the direction of the mean field $\langle \uvec{\omega}\rangle$; (c) the velocity mean fields $\rho_z$ (light red curve, in the direction of $\langle \uvec{\omega}\rangle$) and $\rho_x$ (dark purple curve) as functions of time; (d) the final stationary or rotating distribution of particle velocities (small dots) and the direction of the velocity mean field (large circle); and (e) sample particle trajectories in the final state over $t=100$ time units (thin irregular lines) as well as the full ensemble center of mass trajectory (bold line). Horizontal and vertical dashed lines in panels (a,c) are discussed in the text.}\label{Fig:Examples}
\end{figure}

\subsection{The general case}
\label{Sec:ge}

With $\uvec{z}$-axial symmetry of the distribution $G(\uvec{\omega})$, the matrix $\mathrm{W}$ \eqref{Eq:mW} is diagonal 
and matrix $\mathrm{\Xi}$ \eqref{Eq:mXi} has only the nonvanishing entries $\pm\left\langle\hat\omega_z\right\rangle$. 
Then the matrix determinant \eqref{Eq:DispersionRelation} is a product of two factors so that one of the two equations
\begin{eqnarray}
	0 &=& h_1 + \langle \hat\omega_z^2\rangle h_3-\frac{3}{2K}\;,\label{Eq:Det_z}\\
	0 &=& \left(h_1 + \langle \hat\omega_x^2\rangle h_3-\frac{3}{2K}\right)^2 + \langle \hat\omega_z\rangle^2 h^2_2\;, \label{Eq:Det_xy}
\end{eqnarray}
must hold. If $\langle \hat\omega_z\rangle h_2$ is zero, one can show that no oscillatory instabilities with $\Omega\ne 0$ exist. This includes also the cases discussed in Section~\ref{subsec:Disordered case}. Then the numbers of left and right rotating oscillators around each rotation axis are equal and we can immediately find the solutions with $\Omega=0$ as
\begin{equation}\label{Eq:Kcrit}
	K = \left.\frac{3}{2}\frac{1}{\langle \hat\omega^2 \rangle h_3 + h_1}\right|_{\lambda=2D}\;,
\end{equation}
where $\langle \hat\omega^2 \rangle = \langle \hat\omega_z^2 \rangle$ for Eq.~(\ref{Eq:Det_z}) and $\langle \hat\omega^2 \rangle=\langle \hat\omega_x^2 \rangle=\langle \hat\omega_y^2 \rangle$ for Eq.~(\ref{Eq:Det_xy}). The smaller of these two $K$ values is the critical coupling strength. With a Lorentzian frequency distribution $g(\omega) = \frac{1}{\pi}\frac{\gamma}{(\omega-\omega_0)^2 + \gamma^2}$, the integrals (\ref{Eq:h_1_h_2_h_3}) and $h_3=1/\lambda-h_1$ can 
directly be calculated 
\begin{equation}
\label{Eq:hint}
	h_1 = \frac{\lambda + \gamma}{\left(\lambda+\gamma\right)^2 + \omega_0^2},\quad 
	h_2 = \frac{\omega_0}{(\lambda+\gamma)^2 + \omega_0^2}.
\end{equation}
Inserting these expressions into \eqref{Eq:Det_xy} and \eqref{Eq:Det_z} gives an explicit
formula for the critical coupling strength. To calculate the critical coupling strength in Fig. ~\ref{Fig:Rho_vs_DK}, where $\omega_0=0$ and $\langle \hat\omega^2\rangle=1/3$, we use Eq.~\eqref{Eq:Kcrit}. Equation (\ref{Eq:Kcrit}) with a delta distribution of frequencies, i.e.  $\omega=\omega_0$ and $\gamma=0$, gives the exact same result as in Ref.~\cite{ritort1998solvable} which is thus included in our analysis as a special case.
\subsection{Example: slowly evolving von Mises-Fisher distribution}
\label{subsec:example}

When chiral symmetry is broken, i.e. $\langle \hat\omega_z\rangle h_2 \ne 0$,  oscillatory instabilities can be expected, leading to a partial phase synchrony of the helical trajectories. In this case the swarm center of mass can perform quite regular oscillations whereas individual trajectories appear to be erratic (see Fig.\ref{Fig:Examples}(e2)). Such collective oscillations have recently been observed in dense colonies of E.~coli \cite{chen2017weak}.

As an example shown in Figs.~\ref{Fig:DetSolve},\ref{Fig:Examples}, we study the transition to collective motion in a swarm of globally coupled, self propelled particles of unit velocities $\uvec{v}$ and with helical trajectories. The rotation axes $\uvec{\omega}$ of the particles diffuse and align slowly to their mean direction according to Eq.~\eqref{Eq:FreqDiff} with $d=0.005$ and $\kappa=0.0158$ or $\kappa=0.018$. 
We use these two values to illustrate directed and rotating motions of the particles center of mass.
The frequency distribution $g(\omega)$ is 
Lorentzian with mean frequency $\omega_0=1.0$ and width $\gamma=0.05$.  The coupling strength and the diffusion constant for the velocity vectors are $K=1.38$ and $D=0.2$.

We can apply our linear stability analysis under the assumption of a quasi-static distribution of rotation axes $\uvec{\omega}$. 
We start
with isotropic random initial conditions of uniformly
distributed axes  $\uvec{\omega}$, where the incoherent distribution of velocities $\uvec{\sigma}$ is stable for $K=1.38$. 
As the rotation axes evolve according to \eqref{Eq:FreqDiff}, they 
start to align and $\langle \hat{\omega}_z\rangle$ grows, the moment 
$\langle \omega_z^2\rangle$ is growing and the moments $\langle \omega_x^2\rangle = \langle \omega_y^2 \rangle = 1 -2\langle \omega_z^2\rangle$ are decreasing. With these parameters, the linear stability of the incoherent velocity distribution changes as well. At some point it can become linearly unstable, the velocity vectors start to align and a transition to collective motion is observed.

During the transient we monitor deviation of the rotation axes distribution from the von Mises-Fisher distribution according to Eq. \eqref{Eq:vMF_Distance}. One can see in Fig.~\ref{Fig:Examples}a that systematic deviations are 
smaller than finite ensmble size fluctuations in the equilibrium state, i.e. $\langle \hat\omega_z\rangle$  characterizes the rotation axes distribution completely and we can study the linear stability as a function of $\langle \hat\omega_z\rangle$ alone.

We start with a discussion of linear stability properties of the uniform incoherent
state, according to the analytical expressions of Section~\ref{sec:linear stablity}.
Figure~\ref{Fig:DetSolve}(a,b) shows the critical coupling strength and frequency of unstable modes as a function of $\langle \hat\omega_z\rangle$ according to our 
linear stability analysis (the roots of equation \eqref{Eq:DispersionRelation}
are found numerically). There are two critical branches. One (black) branch corresponds to a transition to a non-oscillating mode, and thus to a directed motion of particles. 
Another (colored) branch corresponds to an oscillating mode, and thus
to center of mass oscillations in the population. We choose the coupling parameter  $K=1.38$, therefore with a gradual increase of $\langle \hat\omega_z\rangle$ the system evolves along a horizontal line in Fig.~\ref{Fig:DetSolve}(a,b). The  first transition at this coupling strength is to a non-oscillating mode at $\langle \hat\omega_z\rangle \approx 0.08$. At  $\langle \hat\omega_z\rangle \approx 0.32$ an oscillating mode also becomes unstable. From the linear analysis we cannot judge, what will be a result of a  competition of these modes. 

Figure \ref{Fig:Examples} shows results of direct numerical simulations, with 
the aim to test the prediction of the linear stability analysis and to explore 
truly nonlinear regimes. We have chosen two values of rotation axes coupling,
$\kappa=0.0158$ in the left column and $\kappa=0.018$ in the right column of Fig. \ref{Fig:Examples}. The difference is that for the former, smaller
value of $\kappa$, the saturation level of   $\langle \hat\omega_z\rangle$
does not exceed the critical value for the oscillatory instability. Thus, here
we expect the directed motion to occur. This is indeed observed in the simulations.
The directed motion itself is illustrated in panel (e1), where one can see that it is superimposed with helical trajectories of the particles. The transition point is, however, delayed in comparison to the theoretical prediction: it happens at time $t\approx 4000$ (see panel (c1)), where the value of    $\langle \hat\omega_z\rangle$ is $0.2$. A delay of bifurcation (compared to the static value $\langle \hat\omega_z\rangle=0.08$) is a general phenomenon for parameter-varying systems, here it might be even enhanced due to finite-size effects.

Another, larger value of $\kappa=0.018$, leads to a saturated level of the alignment of rotation axes at  $\langle \hat\omega_z\rangle \approx 0.5$, which is larger than the second critical value for the instability of the oscillating mode.
Here we observe two transitions, as one can see in panel (c2) of  Fig. \ref{Fig:Examples}. The first transition at $t\approx 1500$ corresponds to the same
value $\langle \hat\omega_z\rangle \approx 0.2$ as in panel (c1). 
In this transition a directed motion with $\rho_z\neq 0$ appears. However, this
motion is a transient episode: it exists only up to time $t\approx 2000$, at which the alignment of frequencies reaches level $\langle \hat\omega_z\rangle \approx 0.38$. Starting from this level, the oscillating mode dominates: a rotation of $\bvec{\rho}$ in the $x$-$y$ plane with $\rho_z\approx 0$ and oscillating values of $\rho_x$ and $\rho_y$ (panel (c2)). The rotational motion of the center of mass is illustrated in panel (e2).

\section{Conclusion}
\label{sec:conclusion}
In conclusion, we have investigated velocity alignment and frequency synchronization 
in a three-dimensional globally coupled 
swarming model with helical trajectories and noise. Unit velocity vectors of the particles precess around individual rotation axes, tend to align into the direction of the mean velocity due to coupling, and are subject to noise. We have derived the condition for the emergence of a non-zero velocity mean field, leading to either a directed motion of the swarm or to collective oscillations. In direct simulations we have only observed  second-order transitions at finite coupling strength, in contrast to a discontinuous transition at infinitesimal small coupling, reported in the singular, deterministic limit  \cite{chandra2019continuous}. A higher order analysis beyond linear stability consideration, such as the multi-scale perturbation method used in the classical Kuramoto model, is still needed to characterize the type and the characteristic exponents of the  synchronization transition.
\begin{acknowledgments}
C.Z. acknowledges the financial support from China Scholarship Council (CSC).
A.P. was supported by the Russian Science Foundation, grant Nr. 17-12-01534.
\end{acknowledgments}
\bibliographystyle{apsrev4-2}
\normalem
\bibliography{3Dnoisykuramoto}

\begin{thebibliography}{18}%
\makeatletter
\providecommand \@ifxundefined [1]{%
 \@ifx{#1\undefined}
}%
\providecommand \@ifnum [1]{%
 \ifnum #1\expandafter \@firstoftwo
 \else \expandafter \@secondoftwo
 \fi
}%
\providecommand \@ifx [1]{%
 \ifx #1\expandafter \@firstoftwo
 \else \expandafter \@secondoftwo
 \fi
}%
\providecommand \natexlab [1]{#1}%
\providecommand \enquote  [1]{``#1''}%
\providecommand \bibnamefont  [1]{#1}%
\providecommand \bibfnamefont [1]{#1}%
\providecommand \citenamefont [1]{#1}%
\providecommand \href@noop [0]{\@secondoftwo}%
\providecommand \href [0]{\begingroup \@sanitize@url \@href}%
\providecommand \@href[1]{\@@startlink{#1}\@@href}%
\providecommand \@@href[1]{\endgroup#1\@@endlink}%
\providecommand \@sanitize@url [0]{\catcode `\\12\catcode `\$12\catcode
  `\&12\catcode `\#12\catcode `\^12\catcode `\_12\catcode `\%12\relax}%
\providecommand \@@startlink[1]{}%
\providecommand \@@endlink[0]{}%
\providecommand \url  [0]{\begingroup\@sanitize@url \@url }%
\providecommand \@url [1]{\endgroup\@href {#1}{\urlprefix }}%
\providecommand \urlprefix  [0]{URL }%
\providecommand \Eprint [0]{\href }%
\providecommand \doibase [0]{https://doi.org/}%
\providecommand \selectlanguage [0]{\@gobble}%
\providecommand \bibinfo  [0]{\@secondoftwo}%
\providecommand \bibfield  [0]{\@secondoftwo}%
\providecommand \translation [1]{[#1]}%
\providecommand \BibitemOpen [0]{}%
\providecommand \bibitemStop [0]{}%
\providecommand \bibitemNoStop [0]{.\EOS\space}%
\providecommand \EOS [0]{\spacefactor3000\relax}%
\providecommand \BibitemShut  [1]{\csname bibitem#1\endcsname}%
\let\auto@bib@innerbib\@empty
\bibitem [{\citenamefont {Lauga}\ and\ \citenamefont
  {Powers}(2009)}]{lauga2009hydrodynamics}%
  \BibitemOpen
  \bibfield  {author} {\bibinfo {author} {\bibfnamefont {E.}~\bibnamefont
  {Lauga}}\ and\ \bibinfo {author} {\bibfnamefont {T.~R.}\ \bibnamefont
  {Powers}},\ }\href@noop {} {\bibfield  {journal} {\bibinfo  {journal}
  {Reports on Progress in Physics}\ }\textbf {\bibinfo {volume} {72}},\
  \bibinfo {pages} {096601} (\bibinfo {year} {2009})}\BibitemShut {NoStop}%
\bibitem [{\citenamefont {Bechinger}\ \emph {et~al.}(2016)\citenamefont
  {Bechinger}, \citenamefont {Di~Leonardo}, \citenamefont {L{\"o}wen},
  \citenamefont {Reichhardt}, \citenamefont {Volpe},\ and\ \citenamefont
  {Volpe}}]{bechinger2016active}%
  \BibitemOpen
  \bibfield  {author} {\bibinfo {author} {\bibfnamefont {C.}~\bibnamefont
  {Bechinger}}, \bibinfo {author} {\bibfnamefont {R.}~\bibnamefont
  {Di~Leonardo}}, \bibinfo {author} {\bibfnamefont {H.}~\bibnamefont
  {L{\"o}wen}}, \bibinfo {author} {\bibfnamefont {C.}~\bibnamefont
  {Reichhardt}}, \bibinfo {author} {\bibfnamefont {G.}~\bibnamefont {Volpe}},\
  and\ \bibinfo {author} {\bibfnamefont {G.}~\bibnamefont {Volpe}},\
  }\href@noop {} {\bibfield  {journal} {\bibinfo  {journal} {Reviews of Modern
  Physics}\ }\textbf {\bibinfo {volume} {88}},\ \bibinfo {pages} {045006}
  (\bibinfo {year} {2016})}\BibitemShut {NoStop}%
\bibitem [{\citenamefont {Tottori}\ \emph {et~al.}(2012)\citenamefont
  {Tottori}, \citenamefont {Zhang}, \citenamefont {Qiu}, \citenamefont
  {Krawczyk}, \citenamefont {Franco-Obreg{\'o}n},\ and\ \citenamefont
  {Nelson}}]{tottori2012magnetic}%
  \BibitemOpen
  \bibfield  {author} {\bibinfo {author} {\bibfnamefont {S.}~\bibnamefont
  {Tottori}}, \bibinfo {author} {\bibfnamefont {L.}~\bibnamefont {Zhang}},
  \bibinfo {author} {\bibfnamefont {F.}~\bibnamefont {Qiu}}, \bibinfo {author}
  {\bibfnamefont {K.~K.}\ \bibnamefont {Krawczyk}}, \bibinfo {author}
  {\bibfnamefont {A.}~\bibnamefont {Franco-Obreg{\'o}n}},\ and\ \bibinfo
  {author} {\bibfnamefont {B.~J.}\ \bibnamefont {Nelson}},\ }\href@noop {}
  {\bibfield  {journal} {\bibinfo  {journal} {Advanced materials}\ }\textbf
  {\bibinfo {volume} {24}},\ \bibinfo {pages} {811} (\bibinfo {year}
  {2012})}\BibitemShut {NoStop}%
\bibitem [{\citenamefont {Xie}\ \emph {et~al.}(2019)\citenamefont {Xie},
  \citenamefont {Sun}, \citenamefont {Fan}, \citenamefont {Lin}, \citenamefont
  {Chen}, \citenamefont {Wang}, \citenamefont {Dong},\ and\ \citenamefont
  {He}}]{xie2019reconfigurable}%
  \BibitemOpen
  \bibfield  {author} {\bibinfo {author} {\bibfnamefont {H.}~\bibnamefont
  {Xie}}, \bibinfo {author} {\bibfnamefont {M.}~\bibnamefont {Sun}}, \bibinfo
  {author} {\bibfnamefont {X.}~\bibnamefont {Fan}}, \bibinfo {author}
  {\bibfnamefont {Z.}~\bibnamefont {Lin}}, \bibinfo {author} {\bibfnamefont
  {W.}~\bibnamefont {Chen}}, \bibinfo {author} {\bibfnamefont {L.}~\bibnamefont
  {Wang}}, \bibinfo {author} {\bibfnamefont {L.}~\bibnamefont {Dong}},\ and\
  \bibinfo {author} {\bibfnamefont {Q.}~\bibnamefont {He}},\ }\href@noop {}
  {\bibfield  {journal} {\bibinfo  {journal} {Sci. Robot}\ }\textbf {\bibinfo
  {volume} {4}} (\bibinfo {year} {2019})}\BibitemShut {NoStop}%
\bibitem [{\citenamefont {Vicsek}\ \emph {et~al.}(1995)\citenamefont {Vicsek},
  \citenamefont {Czir{\'o}k}, \citenamefont {Ben-Jacob}, \citenamefont
  {Cohen},\ and\ \citenamefont {Shochet}}]{vicsek1995novel}%
  \BibitemOpen
  \bibfield  {author} {\bibinfo {author} {\bibfnamefont {T.}~\bibnamefont
  {Vicsek}}, \bibinfo {author} {\bibfnamefont {A.}~\bibnamefont {Czir{\'o}k}},
  \bibinfo {author} {\bibfnamefont {E.}~\bibnamefont {Ben-Jacob}}, \bibinfo
  {author} {\bibfnamefont {I.}~\bibnamefont {Cohen}},\ and\ \bibinfo {author}
  {\bibfnamefont {O.}~\bibnamefont {Shochet}},\ }\href@noop {} {\bibfield
  {journal} {\bibinfo  {journal} {Physical Review Letters}\ }\textbf {\bibinfo
  {volume} {75}},\ \bibinfo {pages} {1226} (\bibinfo {year}
  {1995})}\BibitemShut {NoStop}%
\bibitem [{\citenamefont {Attanasi}\ \emph {et~al.}(2014)\citenamefont
  {Attanasi}, \citenamefont {Cavagna}, \citenamefont {Del~Castello},
  \citenamefont {Giardina}, \citenamefont {Melillo}, \citenamefont {Parisi},
  \citenamefont {Pohl}, \citenamefont {Rossaro}, \citenamefont {Shen},
  \citenamefont {Silvestri} \emph {et~al.}}]{attanasi2014finite}%
  \BibitemOpen
  \bibfield  {author} {\bibinfo {author} {\bibfnamefont {A.}~\bibnamefont
  {Attanasi}}, \bibinfo {author} {\bibfnamefont {A.}~\bibnamefont {Cavagna}},
  \bibinfo {author} {\bibfnamefont {L.}~\bibnamefont {Del~Castello}}, \bibinfo
  {author} {\bibfnamefont {I.}~\bibnamefont {Giardina}}, \bibinfo {author}
  {\bibfnamefont {S.}~\bibnamefont {Melillo}}, \bibinfo {author} {\bibfnamefont
  {L.}~\bibnamefont {Parisi}}, \bibinfo {author} {\bibfnamefont
  {O.}~\bibnamefont {Pohl}}, \bibinfo {author} {\bibfnamefont {B.}~\bibnamefont
  {Rossaro}}, \bibinfo {author} {\bibfnamefont {E.}~\bibnamefont {Shen}},
  \bibinfo {author} {\bibfnamefont {E.}~\bibnamefont {Silvestri}}, \emph
  {et~al.},\ }\href@noop {} {\bibfield  {journal} {\bibinfo  {journal}
  {Physical review letters}\ }\textbf {\bibinfo {volume} {113}},\ \bibinfo
  {pages} {238102} (\bibinfo {year} {2014})}\BibitemShut {NoStop}%
\bibitem [{\citenamefont {Chen}\ \emph {et~al.}(2017)\citenamefont {Chen},
  \citenamefont {Liu}, \citenamefont {Shi}, \citenamefont {Chat{\'e}},\ and\
  \citenamefont {Wu}}]{chen2017weak}%
  \BibitemOpen
  \bibfield  {author} {\bibinfo {author} {\bibfnamefont {C.}~\bibnamefont
  {Chen}}, \bibinfo {author} {\bibfnamefont {S.}~\bibnamefont {Liu}}, \bibinfo
  {author} {\bibfnamefont {X.-q.}\ \bibnamefont {Shi}}, \bibinfo {author}
  {\bibfnamefont {H.}~\bibnamefont {Chat{\'e}}},\ and\ \bibinfo {author}
  {\bibfnamefont {Y.}~\bibnamefont {Wu}},\ }\href@noop {} {\bibfield  {journal}
  {\bibinfo  {journal} {Nature}\ }\textbf {\bibinfo {volume} {542}},\ \bibinfo
  {pages} {210} (\bibinfo {year} {2017})}\BibitemShut {NoStop}%
\bibitem [{\citenamefont {Gr{\'e}goire}\ and\ \citenamefont
  {Chat{\'e}}(2004)}]{gregoire2004onset}%
  \BibitemOpen
  \bibfield  {author} {\bibinfo {author} {\bibfnamefont {G.}~\bibnamefont
  {Gr{\'e}goire}}\ and\ \bibinfo {author} {\bibfnamefont {H.}~\bibnamefont
  {Chat{\'e}}},\ }\href@noop {} {\bibfield  {journal} {\bibinfo  {journal}
  {Physical Review Letters}\ }\textbf {\bibinfo {volume} {92}},\ \bibinfo
  {pages} {025702} (\bibinfo {year} {2004})}\BibitemShut {NoStop}%
\bibitem [{\citenamefont {Chat{\'e}}\ \emph {et~al.}(2008)\citenamefont
  {Chat{\'e}}, \citenamefont {Ginelli}, \citenamefont {Gr{\'e}goire},\ and\
  \citenamefont {Raynaud}}]{chate2008collective}%
  \BibitemOpen
  \bibfield  {author} {\bibinfo {author} {\bibfnamefont {H.}~\bibnamefont
  {Chat{\'e}}}, \bibinfo {author} {\bibfnamefont {F.}~\bibnamefont {Ginelli}},
  \bibinfo {author} {\bibfnamefont {G.}~\bibnamefont {Gr{\'e}goire}},\ and\
  \bibinfo {author} {\bibfnamefont {F.}~\bibnamefont {Raynaud}},\ }\href@noop
  {} {\bibfield  {journal} {\bibinfo  {journal} {Physical Review E}\ }\textbf
  {\bibinfo {volume} {77}},\ \bibinfo {pages} {046113} (\bibinfo {year}
  {2008})}\BibitemShut {NoStop}%
\bibitem [{\citenamefont {Chandra}\ \emph
  {et~al.}(2019{\natexlab{a}})\citenamefont {Chandra}, \citenamefont {Girvan},\
  and\ \citenamefont {Ott}}]{chandra2019continuous}%
  \BibitemOpen
  \bibfield  {author} {\bibinfo {author} {\bibfnamefont {S.}~\bibnamefont
  {Chandra}}, \bibinfo {author} {\bibfnamefont {M.}~\bibnamefont {Girvan}},\
  and\ \bibinfo {author} {\bibfnamefont {E.}~\bibnamefont {Ott}},\ }\href@noop
  {} {\bibfield  {journal} {\bibinfo  {journal} {Physical Review X}\ }\textbf
  {\bibinfo {volume} {9}},\ \bibinfo {pages} {011002} (\bibinfo {year}
  {2019}{\natexlab{a}})}\BibitemShut {NoStop}%
\bibitem [{\citenamefont {Watanabe}\ and\ \citenamefont
  {Strogatz}(1994)}]{watanabe1994constants}%
  \BibitemOpen
  \bibfield  {author} {\bibinfo {author} {\bibfnamefont {S.}~\bibnamefont
  {Watanabe}}\ and\ \bibinfo {author} {\bibfnamefont {S.~H.}\ \bibnamefont
  {Strogatz}},\ }\href@noop {} {\bibfield  {journal} {\bibinfo  {journal}
  {Physica D: Nonlinear Phenomena}\ }\textbf {\bibinfo {volume} {74}},\
  \bibinfo {pages} {197} (\bibinfo {year} {1994})}\BibitemShut {NoStop}%
\bibitem [{\citenamefont {Ott}\ and\ \citenamefont
  {Antonsen}(2008)}]{ott2008low}%
  \BibitemOpen
  \bibfield  {author} {\bibinfo {author} {\bibfnamefont {E.}~\bibnamefont
  {Ott}}\ and\ \bibinfo {author} {\bibfnamefont {T.~M.}\ \bibnamefont
  {Antonsen}},\ }\href@noop {} {\bibfield  {journal} {\bibinfo  {journal}
  {Chaos: An Interdisciplinary Journal of Nonlinear Science}\ }\textbf
  {\bibinfo {volume} {18}},\ \bibinfo {pages} {037113} (\bibinfo {year}
  {2008})}\BibitemShut {NoStop}%
\bibitem [{\citenamefont {Tanaka}(2014)}]{tanaka2014solvable}%
  \BibitemOpen
  \bibfield  {author} {\bibinfo {author} {\bibfnamefont {T.}~\bibnamefont
  {Tanaka}},\ }\href@noop {} {\bibfield  {journal} {\bibinfo  {journal} {New
  Journal of Physics}\ }\textbf {\bibinfo {volume} {16}},\ \bibinfo {pages}
  {023016} (\bibinfo {year} {2014})}\BibitemShut {NoStop}%
\bibitem [{\citenamefont {Chandra}\ \emph
  {et~al.}(2019{\natexlab{b}})\citenamefont {Chandra}, \citenamefont {Girvan},\
  and\ \citenamefont {Ott}}]{chandra2019complexity}%
  \BibitemOpen
  \bibfield  {author} {\bibinfo {author} {\bibfnamefont {S.}~\bibnamefont
  {Chandra}}, \bibinfo {author} {\bibfnamefont {M.}~\bibnamefont {Girvan}},\
  and\ \bibinfo {author} {\bibfnamefont {E.}~\bibnamefont {Ott}},\ }\href@noop
  {} {\bibfield  {journal} {\bibinfo  {journal} {Chaos: An Interdisciplinary
  Journal of Nonlinear Science}\ }\textbf {\bibinfo {volume} {29}},\ \bibinfo
  {pages} {053107} (\bibinfo {year} {2019}{\natexlab{b}})}\BibitemShut
  {NoStop}%
\bibitem [{\citenamefont {Ritort}(1998)}]{ritort1998solvable}%
  \BibitemOpen
  \bibfield  {author} {\bibinfo {author} {\bibfnamefont {F.}~\bibnamefont
  {Ritort}},\ }\href@noop {} {\bibfield  {journal} {\bibinfo  {journal}
  {Physical Review Letters}\ }\textbf {\bibinfo {volume} {80}},\ \bibinfo
  {pages} {6} (\bibinfo {year} {1998})}\BibitemShut {NoStop}%
\bibitem [{\citenamefont {Niedermayer}\ \emph {et~al.}(2008)\citenamefont
  {Niedermayer}, \citenamefont {Eckhardt},\ and\ \citenamefont
  {Lenz}}]{nieder2008}%
  \BibitemOpen
  \bibfield  {author} {\bibinfo {author} {\bibfnamefont {T.}~\bibnamefont
  {Niedermayer}}, \bibinfo {author} {\bibfnamefont {B.}~\bibnamefont
  {Eckhardt}},\ and\ \bibinfo {author} {\bibfnamefont {P.}~\bibnamefont
  {Lenz}},\ }\href@noop {} {\bibfield  {journal} {\bibinfo  {journal} {Chaos:
  An Interdisciplinary Journal of Nonlinear Science}\ }\textbf {\bibinfo
  {volume} {18}},\ \bibinfo {pages} {037128} (\bibinfo {year}
  {2008})}\BibitemShut {NoStop}%
\bibitem [{\citenamefont {Chepizhko}\ and\ \citenamefont
  {Kulinskii}(2010)}]{chepizhko2010relation}%
  \BibitemOpen
  \bibfield  {author} {\bibinfo {author} {\bibfnamefont {A.}~\bibnamefont
  {Chepizhko}}\ and\ \bibinfo {author} {\bibfnamefont {V.}~\bibnamefont
  {Kulinskii}},\ }\href@noop {} {\bibfield  {journal} {\bibinfo  {journal}
  {Physica A: Statistical Mechanics and its Applications}\ }\textbf {\bibinfo
  {volume} {389}},\ \bibinfo {pages} {5347} (\bibinfo {year}
  {2010})}\BibitemShut {NoStop}%
\bibitem [{\citenamefont {Degond}\ \emph {et~al.}(2014)\citenamefont {Degond},
  \citenamefont {Dimarco},\ and\ \citenamefont
  {Mac}}]{degond2014hydrodynamics}%
  \BibitemOpen
  \bibfield  {author} {\bibinfo {author} {\bibfnamefont {P.}~\bibnamefont
  {Degond}}, \bibinfo {author} {\bibfnamefont {G.}~\bibnamefont {Dimarco}},\
  and\ \bibinfo {author} {\bibfnamefont {T.~B.~N.}\ \bibnamefont {Mac}},\
  }\href@noop {} {\bibfield  {journal} {\bibinfo  {journal} {Mathematical
  Models and Methods in Applied Sciences}\ }\textbf {\bibinfo {volume} {24}},\
  \bibinfo {pages} {277} (\bibinfo {year} {2014})}\BibitemShut {NoStop}%
\end{thebibliography}%
\end{document}